\documentclass[10pt,journal]{IEEEtran} 
\usepackage[utf8]{inputenc}
\usepackage[T1]{fontenc}
\usepackage{amssymb,amsmath}
\usepackage{url}
\usepackage[absolute,overlay]{textpos}

\usepackage{pgf}
\usepackage[export]{adjustbox}
\usepackage{bm}
\usepackage{bbm}
\usepackage{booktabs}
\usepackage{url}
\usepackage{cite}
\usepackage{verbatimbox}
\usepackage{enumitem}
\usepackage{amsthm}

\usepackage{colortbl}

\begin{document}
	\title{Polarization Diversity-enabled LOS/NLOS Identification via Carrier Phase Measurements\\
		}		
	\author{Onel L. A. López,~\IEEEmembership{Member,~IEEE,} Dileep Kumar,~\IEEEmembership{Student Member,~IEEE,} 
	and Antti T\"olli,~\IEEEmembership{Senior Member,~IEEE}
	\thanks{Authors are  with  the Centre for Wireless Communications, University of Oulu, Finland, e-mails: \{Onel.AlcarazLopez, Dileep.Kumar, Antti.Tolli\}@oulu.fi}
	\thanks{This research has been supported by the Academy of Finland under Grant 346208 (6G Flagship) and by the Finish Foundation for Technology Promotion.} 
	\thanks{This work has been accepted for publication at IEEE Transactions on Communications. DOI: 10.1109/TCOMM.2023.3236381}
	\thanks{\copyright 2023 IEEE.  Personal use of this material is permitted.  Permission from IEEE must be obtained for all other uses, in any current or future media, including reprinting/republishing this material for advertising or promotional purposes, creating new collective works, for resale or redistribution to servers or lists, or reuse of any copyrighted component of this work in other works.}
	} 	

	\maketitle
\begin{abstract}
The provision of accurate localization is an increasingly important feature of wireless networks. To this end, a reliable distinction between line-of-sight (LOS) and non-LOS (NLOS) radio links is  necessary to avoid degenerative localization estimation biases. Interestingly, LOS and NLOS transmissions affect the polarization of the received signals differently. In this work, we leverage this phenomenon to propose a threshold-based LOS/NLOS classifier exploiting weighted differential carrier phase  measurements over a single link with different polarization configurations. Operation in either full or limited polarization diversity systems is possible. We develop a framework for assessing the performance of the proposed classifier, and show through simulations the performance impact of the reflecting materials in NLOS scenarios. For instance, the classifier is far more efficient in NLOS scenarios with wooden reflectors than in those with metallic reflectors. Numerical results evince the potential performance gains from exploiting full polarization diversity, properly weighting the differential carrier phase  measurements, and using multi-carrier/tone transmissions. Finally, we show that the optimum decision threshold is inversely proportional to the path power gain in dB, while it does not depend significantly on the material of potential NLOS reflectors.
\end{abstract}
	\IEEEpeerreviewmaketitle
\section{Introduction}\label{intro}
Accurate localization (positioning) is a key feature/service of the fifth generation (5G) and beyond wireless networks~\cite{Lima.2020}. 
Applications are numerous, including enhanced emergency call localization, personal navigation, robot/drone/vehicle tracking, autonomous driving, and social networking. Location side-information can also serve, for example, to predict received power levels (including experienced interference) and channel state information (CSI), as a node/beam association criterion, and to enable efficient relay selection and routing strategies, thus facilitating optimized communication network design and operation \cite{Taranto.2014,Lohan.2018,Hu.2020,Lopez.2020}.

The 3rd Generation Partnership Project (3GPP), which is in charge of defining  cellular connectivity standards, started introducing specifications related to cellular-based positioning for 5G in Rel-16 \cite{3GPP.16}.
Later on, meter and sub-meter positioning accuracy targets were set in 3GPP Rel-17 \cite{3GPP.17, tr38-855} (e.g., $0.2$~m and $1$~m positioning accuracy in horizontal and vertical dimension respectively for $90\%$ of the users in industrial Internet of Things use cases).
To support such stringent requirements, operation at millimeter (mm)-wave and terahertz bands, together with the exploitation of large bandwidths and highly directional transmissions from massive antenna arrays, are usually required since they inherently provide higher temporal, spatial, and angular resolutions \cite{3GPP.17, tr38-855,Lin.2018,Dwivedi.2021}. 
These techniques and operation regimes are leveraged by radio access technology (RAT)-dependent positioning solutions, such as Time-of-Arrival~(ToA), Time-Difference-of-Arrival (TDoA), Multi-Cell Round-Trip-Time (Multi-RTT), Angle-of-Departure (AoD), Angle-of-Arrival (AoA), and Enhanced Cell ID (E-CID), supported by 5G, in addition to RAT-independent solutions relying on  Global Navigation Satellite System (GNSS), barometric pressure, and other techniques~\cite{Lima.2021}. However, the current positioning methods alone cannot meet the aforementioned requirements, as shown by the evaluation results in~\cite{Samsung19, Ericsson19, inteEval19}. Thus, to support sub-meter level accuracy, it was recently agreed to consider the carrier-phase based ranging as one of the candidates for the study of RAT-dependent high-accuracy positioning \cite{Fraunhofer19,CATT, DanKookUniversity}.
\subsection{The LOS/NLOS Identification Problem \& State-of-the-Art Solutions}\label{stateArt}
Traditionally, scattering or multi-path propagation has been considered a critical problem for attaining accurate positioning. Consequently, multi-path mitigation techniques have been continuously developed to combat its detrimental effect. More recently, multi-path exploitation mechanisms, where scattered and reflected signals are not seen as a disturbance but as a source of additional information \cite{Witrisal.2016,Wymeersch.2021}, are also proposed for greater performance enhancements. In either case, one problem that persists lies in determining whether transmitter and receiver are, or are not, under line-of-sight (LOS). 
Note that in many environments there may be obstacles blocking the LOS connection between the nodes (i.e., non-LOS (NLOS) situation), thereby potentially introducing biases in the distance estimation. Therefore, distinguishing NLOS from LOS channels is critical for attaining accurate positioning estimates, and constitutes the focus of our work. 
 
In recent years, the LOS/NLOS identification problem has received significant attention. Basically, LOS/NLOS classifiers are based on $i$) thresholding, where a relevant (or combination of relevant) statistic(s) is compared to a given threshold to infer a decision, e.g., \cite{Borras.1998,Heidari.2008,Kegen.2013,Chenshu.2015,Silva.2020, Zhu.2019, Sosnin.2021}, or $ii$) machine learning (ML), where the classification is transparently made, and only the training and/or intelligence mechanism needs to be designed, e.g., \cite{Xiao.2015, Liu.2019, Wu.2018, Bregar.2018, Park.2020, Vales.2020, Huang.2020,Fan.2019, Kirmaz.2021,Huang.2022}.
In general, statistics/features such as root mean square delay spread (RMS-DS), maximum amplitude, rise time, received signal strength (RSS), and other similar metrics (mainly extracted from the channel impulse response (CIR)) are usually exploited.

In the case of threshold-based classifiers, the authors in \cite{Borras.1998} modeled LOS and NLOS hypothesis as being
corrupted by additive noise with different probability distributions. They solved the binary hypothesis test under several assumptions. 
Similarly, Bayesian likelihood functions are also utilized in \cite{Heidari.2008,Kegen.2013} for LOS/NLOS identification exploiting the RSS, ToA, and/or RMS-DS statistics. 
Meanwhile, multi-carrier phase measurements  are leveraged in \cite{Chenshu.2015} for LOS/NLOS identification in WiFi networks using the so-called PhaseU framework. Specifically, the relevant statistic for the hypothesis testing consists of the weighted sum of the variance of the phases measured over multiple frequency transmissions. More recently, the authors in \cite{Silva.2020, Zhu.2019} exploited distance estimate residuals from multiple TRPs, however, 
as in \cite{Wu.2018}, LOS from several TRPs is still a requirement. The average NLOS channel power per subcarrier is leveraged in \cite{Sosnin.2021} to perform LOS/NLOS link classification. Therein, NLOS links are discarded, and a positioning decision is made considering only the measurements from those TRPs that establish LOS links with the user equipment (UE). However, the authors in \cite{Sosnin.2021} may have implicitly assumed operation at regular microwave frequency bands since NLOS links are considered as multi-tap with no dominant tap. Then, given the use of a potentially massive number of antennas when operating at mm-wave or higher frequencies, NLOS links may exhibit a single or dominant tap, thus, their proposed framework cannot be, at least straightforwardly, applied in such a case.

In the case of ML-based classifiers, RSS measurements from WiFi signals are  exploited in \cite{Xiao.2015}, while in \cite{Liu.2019} the training data and testing data sets are generated by a GNSS software receiver  using intermediate frequency signal collected from an indoor pseudolite system. The spatial measurement diversity from multiple transmission reception points (TRPs) operating with TDOA is leveraged in \cite{Wu.2018}. However, LOS from several TRPs is required for the proposed method to  work effectively. Meanwhile, ultra-wideband (UWB) systems are the focus in \cite{Bregar.2018, Park.2020}. CIR realizations are used for training a deep neural network (NN) in \cite{Bregar.2018}, while a hybrid deep learning and transfer learning method is proposed in \cite{Park.2020}. A downsampled power delay profile, as a simpler alternative to the CIR, is proposed in \cite{Vales.2020} to feed a low-complexity deep NN. Moreover, three ML classifiers are considered in \cite{Huang.2020}, namely, least-square support vector machine, random forest, and NN, which are selectively fed with channel features extracted from data measurements such as the minimum received power over delay samples, kurtosis and skewness of the received power, rise time, RMS-DS, Rician factor, and angular-related features. The main conclusion here is that the selection of the training features, and not the ML classifier itself, is the most crucial factor in the performance of LOS identification. Different from the previous works,  unsupervised ML is proposed in \cite{Fan.2019, Kirmaz.2021}. Specifically, an expectation maximization mechanism for Gaussian mixture models is developed in \cite{Fan.2019}. Therein, essential features of a UWB received signal, i.e., number of channel paths, mean excess delay, and RMS-DS, are used as input to the ML classifier. Meanwhile, a channel feature selection process is introduced in \cite{Kirmaz.2021} to select only useful channel features for the classification. We encourage the readers to refer to \cite{Huang.2022} for further details about ML-enabled channel modeling/prediction, and specifically LOS/NLOS identification, including an exhaustive state-of-the-art literature revision and discussions on relevant performance insights.

Notice that the main disadvantages of ML mechanisms compared to threshold-based classifiers are related to their data/time-hungry characteristics, complexity, and/or difficulty to probabilistically interpret the outputs. This may make such approaches unaffordable in many practical scenarios, which motivates our current work to focus on a simple threshold-based classifier.
\subsection{Contributions and Organization of the Paper}\label{contr}
As discussed above, we propose a simple threshold-based LOS/NLOS classifier in this work.  
Specifically, we consider a single pair of TRPs connected through a single path, therefore, the threshold-based mechanisms proposed in \cite{Kegen.2013,Wu.2018, Silva.2020, Zhu.2019, Sosnin.2021} cannot be applied here. Moreover, while most of the approaches discussed in Section~\ref{stateArt} are data/time-hungry and/or require sampling the received signals' features over many time instances to compute relevant decision statistics, our proposed framework is simpler as it relies on at most four instantaneous measurements, thus, it is more suitable for low-latency systems.
Specifically, our proposed classifier exploits an interesting physical phenomenon, namely, the polarization of received signals is affected differently in LOS and NLOS conditions \cite{Kwon.2011}. To the best of our knowledge, our proposal is the first to exploit instantaneous carrier phase measurements and polarization-diversity.
Our specific contributions with respect to the state-of-the-art research are four-fold:
\begin{itemize}
    \item We propose a threshold-based LOS/NLOS classifier exploiting transmit/receive polarization diversity. Specifically, an access point (AP) transmits a tone/carrier whose phase is measured at the UE over at most four transmit-receive polarization configurations. The  relevant decision metric is conformed by weighted differential phase measurements. A LOS/NLOS decision is adopted after comparing such relevant metric  to a decision threshold. Observe that our approach is substantially different from PhaseU in \cite{Chenshu.2015}, where the variance of the phase measurements, thus, large sampling-overhead, is required, and polarization diversity is not exploited.
    \item We present several weighting approaches for the differential phase measurements. Specifically, the equal (EQU) weighting approach is noise-agnostic  as it assigns equal weights to the differential phase measurements, while the more advanced minimum noise variance (MNV) and noise variance -proportional (NVP) weighting approaches exploit noise statistics, and thus perform better. Their specific pros and cons are discussed and numerically validated, while the performance of PhaseU \cite{Chenshu.2015} is also assessed for benchmarking. It is shown that the  NVP weighting approach stands out among all weighting configurations considered while potentially outperforming the PhaseU approach in scenarios with relatively high path power gain and/or when using multiple tones.
    \item We develop a framework for assessing the performance of the proposed LOS/NLOS classifier under different polarization-diversity and scenario configurations. Such a framework includes a detailed characterization of the influence of LOS/NLOS propagation, AP and UE orientation, transmit/receive polarization, reflectors' electric properties, and signal strength on the carrier phase measurements. Our proposal can be applied to limited polarization-diversity systems (i.e., dual polarization at either transmitter or receiver);
    \item The simulation results reveal that: $i$) significant performance gains can be attained from properly weighting, e.g., via NVP, the differential carrier phase  measurements, and exploiting multi-carrier/tone transmissions; $ii$) the performance of the proposed classifier depends critically on the NLOS environment (e.g., being seriously affected in scenarios with metallic reflectors); $iii$) the optimum decision threshold in terms of average error rate (AER) remains approximately the same independently of the material of potential NLOS reflectors, while it decreases following a power-law as the average path/link power gain increases; and $iv$) an inappropriate weighting of differential carrier phase  measurements in a full polarization-diversity system, and also operating with limited diversity, degrades the classifier performance significantly. 
\end{itemize}

The remainder of this paper is organized as follows. Section~\ref{system} introduces the system model and the problem of interest. In Section~\ref{prop}, we characterize the propagation conditions under LOS and NLOS, and model the carrier phase measurements. Section~\ref{hypT} presents the proposed polarization diversity-enabled LOS/NLOS classifier exploiting carrier phase measurements. Finally, Section~\ref{results} presents numerical results, and Section~\ref{conclusions} concludes the article. 

\textbf{Notation:} Boldface lowercase/uppercase letters denote column vectors/matrices, with the only exception of $\mathbf{F}_u$, which denotes a two-element field pattern vector by convention. Superscript $(\cdot)^T$ denotes the transpose operation, while $\lVert\cdot\rVert$ is the Euclidean norm of a vector, and $\lVert\cdot\rVert_F$ is the Frobenius norm of a matrix. $\mathbb{C}$, $\mathbb{R}$
are respectively the set of complex and real numbers.
Additionally, $\Pr(A)$ is the probability of occurrence of event $A$, $\Pr(A|B)$ is the conditional probability of occurrence of event $A$ given the occurrence of event $B$, $\min\{\cdot\}$ is the minimum function, $\mathrm{mod}(\cdot,\cdot)$ is the modulo operation, while $\Re\{\cdot\}$ and $\Im\{\cdot\}$ are respectively the real and imaginary part operators.
Finally, $\mathrm{arctan2}(\cdot, \cdot)$ is the four-quadrant inverse tangent, which is given by
\begin{align}
    \mathrm{arctan2}(y,x)=2\arctan\frac{y}{\sqrt{x^2+y^2}+x}.
\end{align}
\section{System Model}\label{system}
\begin{figure}[t!]
	\centering
	\includegraphics[width=0.45\textwidth]{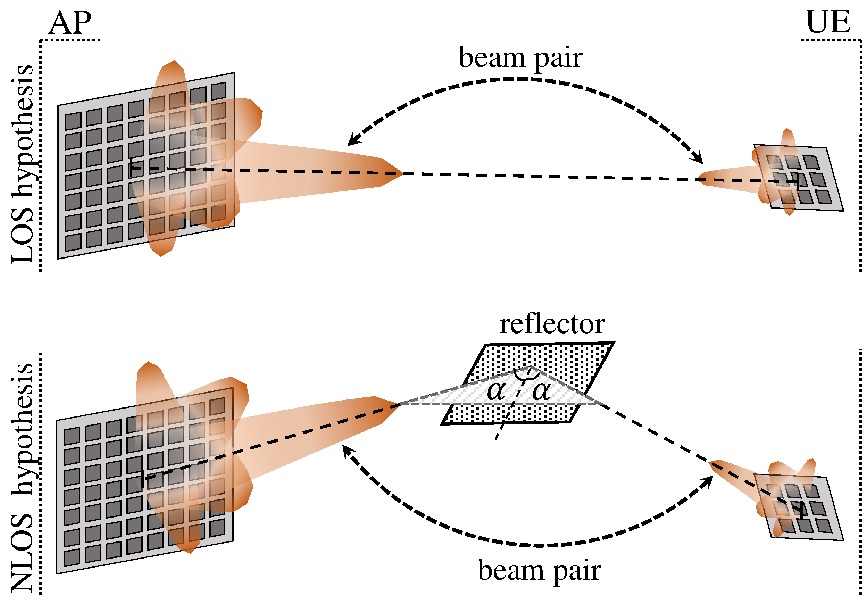}
	\caption{System model: a) paired beams of AP and UE under LOS (top), and b) NLOS (bottom) conditions.}
	\label{Fig1}
	\vspace{-3mm}
\end{figure}

Consider the system model illustrated in Fig.~\ref{Fig1}, where an AP, and a UE at an unknown position, have paired their main transmit/receive (mm-wave multi-antenna) beams.
Beam management techniques, including beam establishment, beam refinement, and beam tracking procedures, are usually exploited for this \cite{Giordani.2019,Bang.2021}. Assume a single path, which can be either LOS (Fig.~\ref{Fig1}a) or NLOS (Fig.~\ref{Fig1}b) motivated by the use of massive transmit antenna arrays and mm-wave spectrum.\footnote{In the case that the AP and UE are communicating through several beam pairs (single paths), one can still straightforwardly leverage our proposed method to classify them as LOS/NLOS.}  Moreover, the scattering phenomenon is ignored, although its effect could be included in the noise modeling introduced later in Section~\ref{carrier}. In fact, some numerical results on this are discussed in Section~\ref{Scatt}. 
\subsection{Channel Model and Polarization Configuration}
We assume that both AP and UE are equipped with collocated vertically/horizontally-polarized antenna elements.
Let $\mathbf{F}_t$ and $\mathbf{F}_r$ be the field pattern of the transmit and receive antennas, which respectively depend on the tuples $(\phi^t,\theta^t)$ and $(\phi^r,\theta^r)$. Here, $\phi^t,\phi^r\in[-\pi,\pi]$ are the departure and arrival azimuth angles, while $\theta^t,\theta^r\in[-\pi/2,\pi/2]$ are the departure and arrival elevation angles. The field patterns are two-element vectors containing the vertically and horizontally polarized component of the antenna array response as
\begin{align}
  \mathbf{F}_u=\begin{bmatrix}
    F_u^V(\phi^u,\theta^u)\\
    F_u^H(\phi^u,\theta^u)
  \end{bmatrix},\qquad u\in\{t,r\},
  \label{F}
\end{align}
and they capture the transmit/receive beamforming effects.

The AP transmits a constant tone\footnote{In practice, the continuous wave  may originate directly from the local oscillator signal, or can be obtained by a FSK-modulation with a constant input (all $1$s or $0$s) \cite{Zand.2019} as in the constant tone extension signal from Bluetooth Core Specification 5.2 \cite{blu.2019}.} at frequency $f$, thus, the electric field
experienced by the receiver (at the output of the receive beamformer) can be written as
\begin{align}
E_r &= \sqrt{PZ_0}\mathbf{F}_r^T\mathbf{M}\mathbf{F}_t\exp{\!\left(-j\frac{2\pi}{\lambda}d\right)}= \zeta \mathbf{F}_r^T \mathbf{M}\mathbf{F}_t,\label{Er}
\end{align}
\begin{figure*}[t!]
	\begin{align}
		\mathbf{Q}(\beta_x,\beta_y,\beta_z) =&
		\underbrace{
			\begin{bmatrix}
				\cos (-\beta_z) & -\sin (-\beta_z) & 0\\
				\sin (-\beta_z) & \cos (-\beta_z) & 0\\
				0 & 0 & 1
		\end{bmatrix}}_{\text{rotation matrix about\ }z}
		\underbrace{
			\begin{bmatrix}
				\cos (-\beta_y) & 0 & \sin (-\beta_y)\\
				0 & 1 & 0\\
				-\sin (-\beta_y) & 0 & \cos (-\beta_y)
		\end{bmatrix}}_{\text{rotation matrix about\ }y}\underbrace{
			\begin{bmatrix}
				1 & 0 & 0\\
				0 & \cos (-\beta_x) & -\sin(-\beta_x)\\
				0 & \sin (-\beta_x) & \cos (-\beta_x)
		\end{bmatrix}}_{\text{rotation matrix about\ }x}\nonumber\\
		=&
		\begin{bmatrix}
			\cos \beta_z & \sin \beta_z & 0\\
			-\sin \beta_z & \cos \beta_z & 0\\
			0 & 0 & 1
		\end{bmatrix}
		\begin{bmatrix}
			\cos \beta_y & 0 & -\sin \beta_y\\
			0 & 1 & 0\\
			\sin \beta_y & 0 & \cos \beta_y
		\end{bmatrix}
		\begin{bmatrix}
			1 & 0 & 0\\
			0 & \cos \beta_x & \sin\beta_x\\
			0 & -\sin \beta_x & \cos \beta_x
		\end{bmatrix}
		\nonumber\\
		=&\begin{bmatrix}
			\cos \beta_z\cos \beta_y & \sin \beta_z\cos\beta_x+\cos \beta_z\sin\beta_y\sin\beta_x & \sin \beta_z\sin\beta_x-\cos \beta_z\sin\beta_y\cos\beta_x\\
			-\sin \beta_z\cos \beta_y & \cos \beta_z\cos\beta_x-\sin \beta_z\sin\beta_y\sin\beta_x & \cos \beta_z\sin\beta_x+\sin \beta_z\sin\beta_y\cos\beta_x\\
			\sin \beta_y & -\cos \beta_y\sin\beta_x & \cos\beta_y\cos\beta_x
		\end{bmatrix}\tag{5}\label{Q}
	\end{align}\hrule
\end{figure*}

\vspace{-3mm}
\noindent where $P$ is the power gain of the path, $Z_0=120\pi$ is the characteristic impedance of vacuum, $\lambda=c/f$ is the wavelength being $c=3\times 10^8$ m/s the speed of light, $d$ is the distance traveled by the signal, 
 $\zeta\triangleq \sqrt{PZ_0}\exp{(\!-j2\pi d/\lambda)}$, and $\mathbf{M}$ is the $2 \!\times\! 2$ polarization coupling matrix, which describes how the path affects the polarization configuration, i.e., how the polarization changes on the way from the AP to the UE. Commonly, $\mathbf{M}$ is modeled using the  horizontal/vertical cross-polarization ratio metrics together with random coefficients \cite{Kwon.2011}. More recently, it was shown in  \cite{Jaeckel.2012} that $\mathbf{M}$ can be modeled as a Jones matrix in the case of geometrical channels, e.g., mm-wave channels. Therefore, we adopt this latter framework in Section~\ref{prop} to capture the polarization changes due to the signal propagation.
 	
 	Without loss of generality, we set the AP's antenna orientation axes as reference. The AP's antenna array lies in the $yz-$plane. Then, the 3D orientation of the UE's antenna array is defined by the Euler angles $\beta_x,\beta_y,\beta_z\in[-\pi/2,\pi/2]$ radians,  which correspond to the angular rotation  about axes $x$, $y$, $z$  needed to match the orientation of the AP's antenna array. Moreover, the AP transmits with either a vertical
or horizontal polarization at a time. 
\subsection{Goal \& Hypothesis}\label{goalH}
In this work, we are concerned with the LOS/NLOS identification problem for a given single path (beam pair). Specifically, the goal is to determine whether: 
\begin{itemize}
	\item[$H_0$:] the beams are paired through a direct (LOS) path as in Fig.~\ref{Fig1}a; or
	\item[$H_1$:] the beams are paired through a reflected (NLOS) path as in Fig.~\ref{Fig1}b. 
\end{itemize}

Note that both the phase and amplitude of the incident polarized signal components are affected differently depending on whether the channel is LOS or NLOS, which can be exploited for LOS/NLOS identification. In the following, we rely on the geometrical theory  for polarized channels \cite{Kwon.2011,Jaeckel.2012} to illustrate this, and specifically discuss how phase measurements can be leveraged.
\section{Propagation and Phase Measurements}\label{prop}
\subsection{LOS Propagation}\label{LOS}
It is shown in \cite{Jaeckel.2012} that the polarization coupling matrix under LOS conditions has the form of a rotation matrix with rotation angle $\vartheta$, thus, 
\begin{align}
\mathbf{M} = \begin{bmatrix}
\cos\vartheta& \sin\vartheta\\
-\sin\vartheta & \cos\vartheta
\end{bmatrix}.\label{M}
\end{align}
\begin{figure*}[t!]
	\centering
	\includegraphics[width=0.94\textwidth]{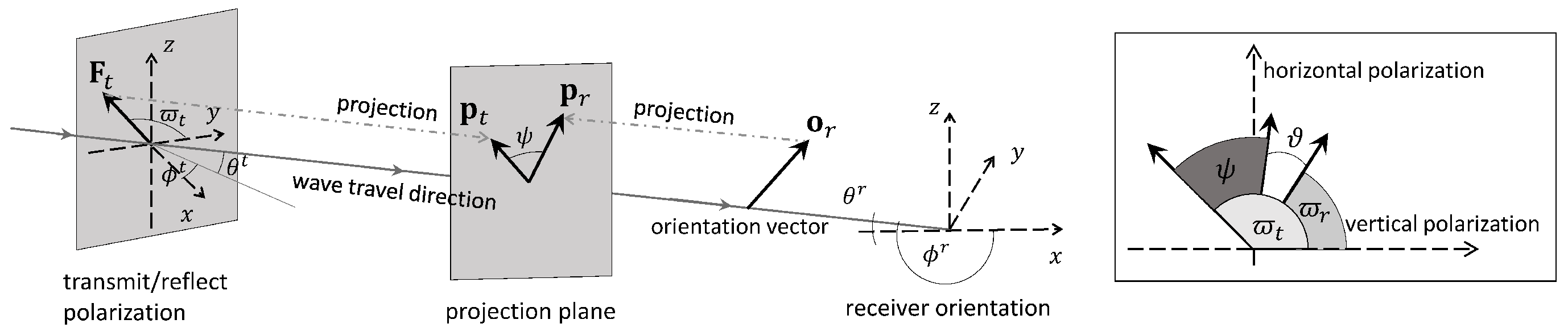}
	\caption{Illustration of the geometric polarization \cite{Jaeckel.2012}.}
	\label{Fig2}
\end{figure*}

Next, we illustrate how $\vartheta$ can be obtained. First, let us define $\mathbf{Q}\in\mathbb{R}^{3\times 3}$ as shown in \eqref{Q} at the top of the page to describe the axis orientation change of the receive antenna (which lies in its local $xz$ plane) with respect to the transmit antenna in 3D space. Let $\mathbf{o}_r$ be the receive orientation vector (e.g., $[1,0,0]^T$ and $[0,1,0]^T$ for vertically and horizontally-polarized receptions, respectively). Then, the transformed orientation vector in global coordinates is given by
\begin{align}
	\setcounter{equation}{5}
\bar{\mathbf{o}}_r = \mathbf{Q}\mathbf{o}_r.
\end{align}
Now, we rotate the orientation vector by $p=-\phi^r-\pi/2$ in the azimuth direction and $q=\theta^r$ in the elevation direction to
match the orientation of the transmitter (thus, the wave travel direction lies now in the $y-$direction) as
\begin{align}
\tilde{\mathbf{o}}_r &= \begin{bmatrix}
\cos p& -\sin p & 0\\
\cos q  \sin p & \cos q \cos p & -\sin q\\
\sin q\sin p & \sin q \cos p & \cos q
\end{bmatrix}\bar{\mathbf{o}}_r\nonumber\\
&=-\begin{bmatrix}
\sin \phi^r\!\!& -\cos \phi^r \!\!& 0\\
\cos \theta^r  \cos \phi^r \!\!& \cos \theta^r \sin \phi^r \!\!& \sin \theta^r\\
\sin \theta^r\cos \phi^r \!\!& \sin \theta^r \sin \phi^r \!\!& -\!\cos \theta^r
\end{bmatrix}\!\bar{\mathbf{o}}_r.\label{o}
\end{align}
Since the projection plane lies in the $xz$ plane due to the placement of the transmitter, we simply omit the $y-$component of $\tilde{\mathbf{o}}_r$, and calculate the projection of the orientation vector on the projection plane as
\begin{align}
\mathbf{p}_r=\frac{\mathbf{A}_{xz}\tilde{\mathbf{o}}_r}{\rVert\mathbf{A}_{xz}\tilde{\mathbf{o}}_r\lVert},
\end{align}
where $\mathbf{A}_{xz}=\begin{bmatrix}
1 & 0 & 0\\
0 & 0 & 1
\end{bmatrix}$. Meanwhile, the transmit polarization vector is given by
\begin{align}
\mathbf{p}_t = \frac{\mathbf{F}_t}{\rVert\mathbf{F}_t\lVert}, 
\end{align}
while the angle between transmit and receive polarization vectors obeys
\begin{align}
\psi = \text{arccos}(\mathbf{p}_t^T\mathbf{p}_r).
\end{align}
The difference between angles 
\begin{align}
\varpi_t &= \text{arctan2}\big(F_t^H(\phi^r,\theta^r),F_t^V(\phi^t,\theta^t)\big),\\
\varpi_r &= \text{arctan2}\big(F_r^H(\phi^r,\theta^r),F_r^V(\phi^t,\theta^t)\big)
\end{align}
represents the polarization mismatch between the receive and transmit antenna if both were aligned on the same optical axis. Note that $\psi$,  however, takes the different orientation of the receive antenna into account. Finally, one obtains 
\begin{align}
\vartheta = \varpi_t-\varpi_r-\psi.\label{var}
\end{align}
All the above process is illustrated geometrically in Fig.~\ref{Fig2}.
\subsection{NLOS Propagation}\label{NLOS}
Assume for simplicity a single-reflection path in the NLOS scenario. The reflecting surface is rotated about axes $x$, $y$, $z$ by $\delta_x,\delta_y,\delta_z\in[-\pi/2,\pi/2]$ radians (Euler angles) with respect to the AP antennas. Then, under the NLOS hypothesis, the polarization coupling matrix has the form
\begin{align}
\mathbf{M}=\underbrace{\begin{bmatrix}
\cos\vartheta_2 & \sin\vartheta_2\\
-\sin\vartheta_2 & \cos\vartheta_2
\end{bmatrix}}_{\scriptsize\begin{tabular}{c}
rotation matrix:\\ reflector$-$UE
\end{tabular}} \underbrace{{\begin{bmatrix} 
	R_{\perp} & 0\\
	0 & R_{\parallel}
	\end{bmatrix}}}_{\text{reflection matrix}}
\underbrace{\begin{bmatrix}
	\cos\vartheta_1 & \sin\vartheta_1\\
	-\sin\vartheta_1 & \cos\vartheta_1
	\end{bmatrix}}_{\scriptsize\begin{tabular}{c}
	rotation matrix:\\ AP$-$reflector
	\end{tabular}},\label{Mnlos}
\end{align}
where  $\vartheta_1$ and $\vartheta_2$ are the rotation angles for the incident and reflected electromagnetic fields, respectively. Similarly, $R_{\parallel}$ and $R_{\perp}$ are the parallel  and perpendicular reflection coefficients of the reflector, respectively. The reflection coefficients are defined as \cite{Ma.2018}
\begin{align}
R_{\perp} &= \frac{\cos\alpha-\sqrt{\varepsilon-\sin^2\alpha}}{\cos\alpha+\sqrt{\varepsilon-\sin^2\alpha}}, \label{Rper} \\
R_{\parallel} &=\frac{\varepsilon\cos\alpha-\sqrt{\varepsilon-\sin^2\alpha}}{\varepsilon\cos\alpha+\sqrt{\varepsilon-\sin^2\alpha}},\label{Rpar}
\end{align}
where $\varepsilon = \varepsilon_{r}-j60\kappa\lambda$, $\alpha$ is the angle of incidence as illustrated in Fig.~\ref{Fig1}, and $\varepsilon_{r}$ and $\kappa$ are the normalized relative dielectric constant and the conductivity of the reflecting surface at the operation frequency, respectively.\footnote{At sufficiently high frequencies, even a metal behaves as a dielectric \cite{Sverre.2015}.} Observe that NLOS converges to LOS  for $\alpha=\pi/2$, for which we can write \eqref{Mnlos} as \eqref{M} with $\vartheta=\vartheta_1+\vartheta_2$. 

Table~\ref{table1} compiles the electric properties of some materials that appear commonly as reflectors in wireless communication networks. Their corresponding parallel and perpendicular reflection coefficients are respectively plotted in Fig.~\ref{Fig3} and Fig.~\ref{Fig4} as a function of the angle of incidence $\alpha$. Note that $\alpha$ depends not only on the transmit antenna, receive antenna, and reflector orientations, but also on their corresponding positions. Later, in Section~\ref{results}, we generate random values for $\alpha$ to account for different node deployment scenarios.
\begin{table}
	\centering
	\caption{Electric properties of the considered reflectors \cite{Sverre.2015,Aleksandar.1998,Alekseev.2000,Wu.2015}.}
	\begin{tabular}{p{4cm}|p{1.6cm}|p{1.6cm}}
		\hline
		Materials & $\epsilon_r$ & $\kappa$ (S/m) \\ \midrule
		glass   &  $6$ & $10^{-14}$ \\
		wood  &  $1.2$ & $10^{-4}$ \\		
		moist concrete & $2.3$ & $10^{-2}$ \\
		distilled water   & 4 &  5\\
		conductor (metal-like/alloys)   & $30$ &  $10$\\	
		\bottomrule
	\end{tabular}\label{table1}
\end{table}

The rotation angles $\vartheta_1$ and $\vartheta_2$ in \eqref{Mnlos} can be obtained by treating the reflecting surface as a virtual antenna and exploiting the framework in Section~\ref{LOS}. By doing this, Fig.~\ref{Fig2} can be re-utilized to visualize/understand the polarization changes undergone over each, the incident and reflected, path component.
\subsubsection{On the computation of $\vartheta_1$}
The reflecting surface is modeled as a virtual receive antenna with arrival azimuth and elevation angles $\alpha$ and $\theta_t-\delta_x$, respectively. Then, $\vartheta_1$ can be obtained by following the procedure for computing \eqref{var} but using $\mathbf{Q}(\delta_x,\delta_y,\delta_z)$ when evaluating \eqref{Q}, and setting $\phi_r\leftarrow \alpha$ and $\theta_r\leftarrow \theta_t-\delta_x$.
\subsubsection{On the computation of $\vartheta_2$}
The reflecting surface is modeled as a virtual transmit antenna. Then, $\vartheta_2$ can be obtained by following the procedure for computing \eqref{var} but using $\mathbf{Q}(\beta_x-\delta_x,\beta_y-\delta_y,\beta_z-\delta_z)$ when evaluating \eqref{Q}.
\begin{figure}[t!]
	\centering
	\qquad	\includegraphics[width=0.46\textwidth]{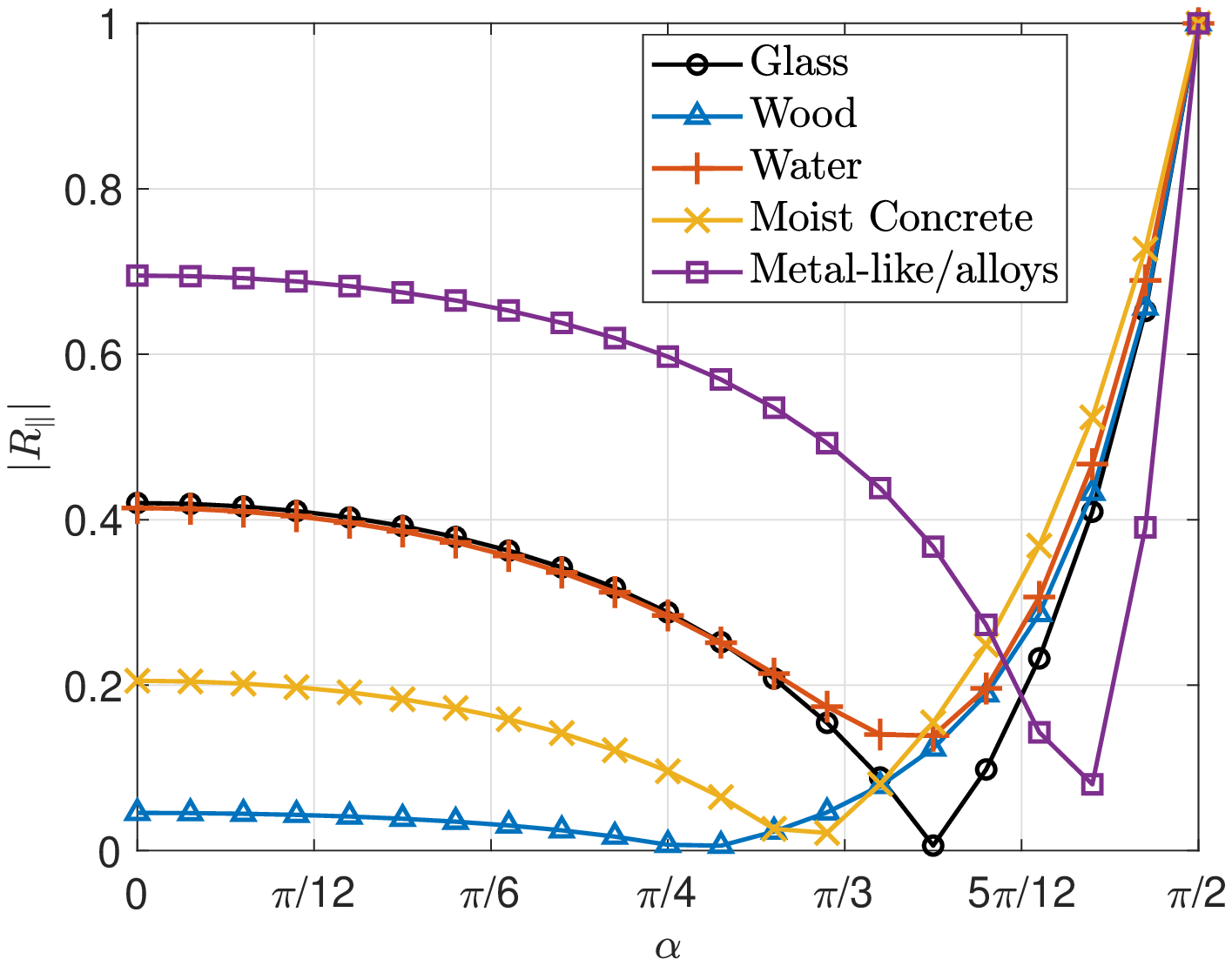}\\
	\includegraphics[width=0.46\textwidth]{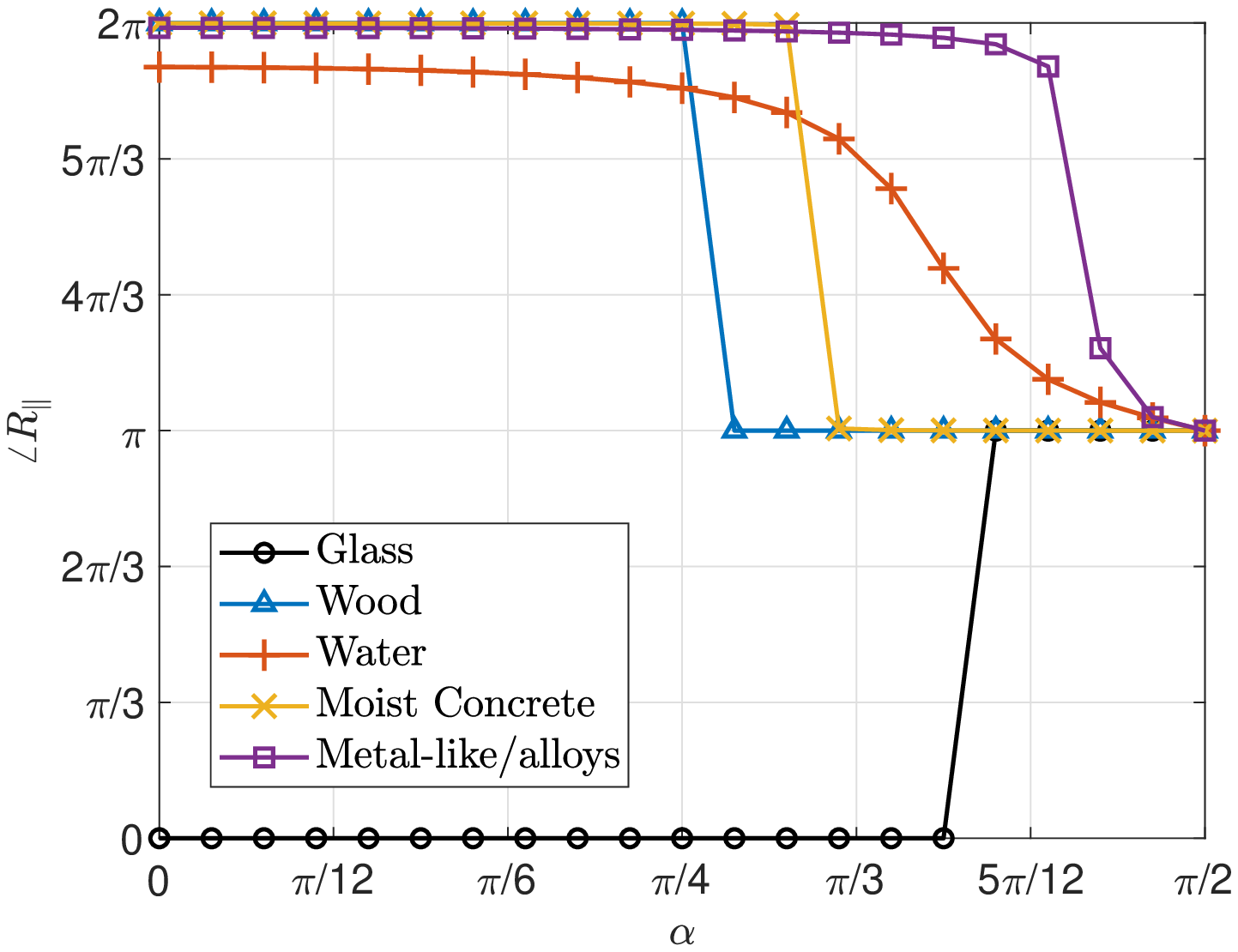} 
	\caption{Parallel reflection coefficient of the materials in Table~\ref{table1} as a function of incidence angle $\alpha$ and specified by amplitude (top) and phase (bottom).}
	\label{Fig3}
\end{figure}
\begin{figure}[t!]
	\centering
	\qquad \includegraphics[width=0.46\textwidth]{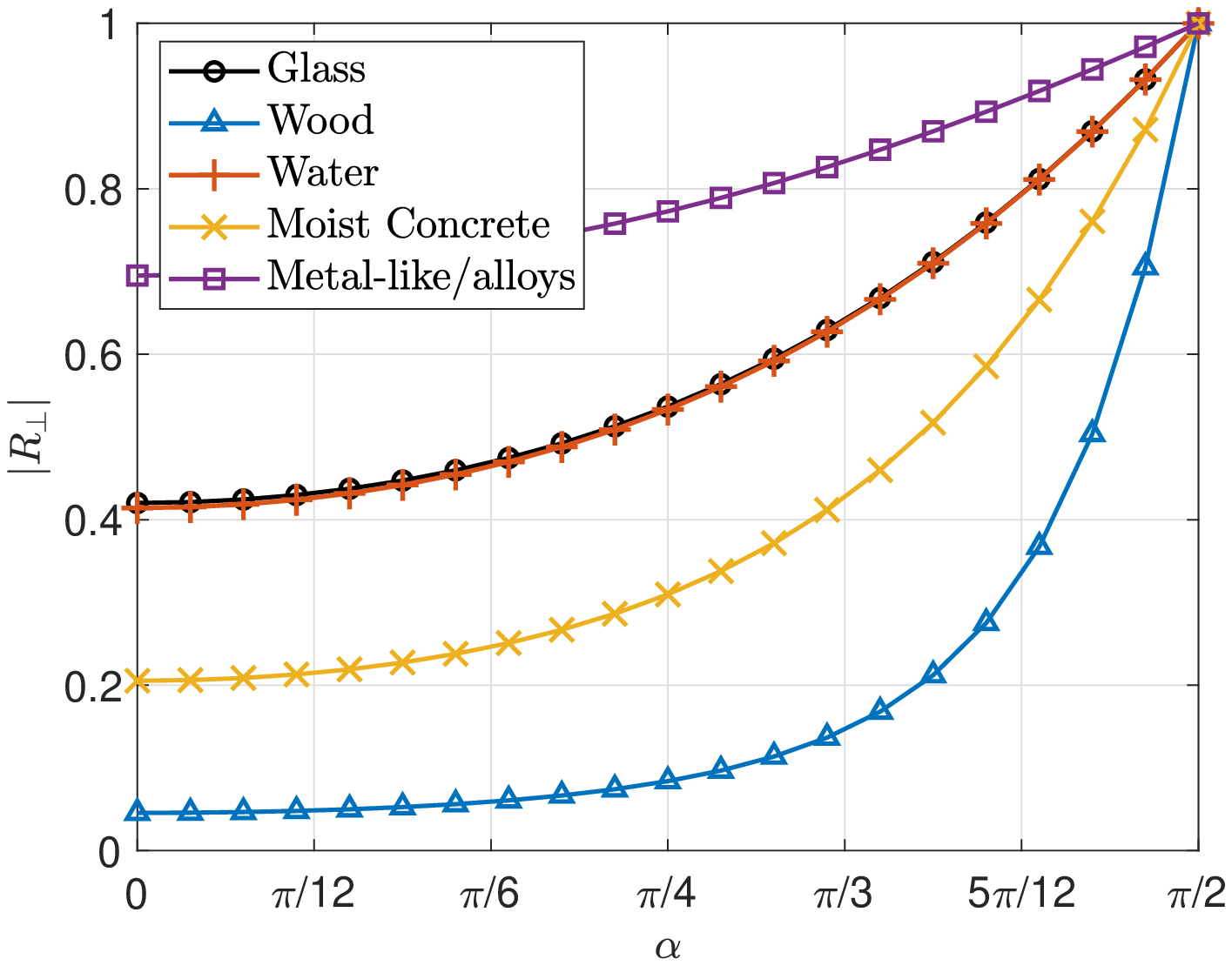}\\
	\includegraphics[width=0.46\textwidth]{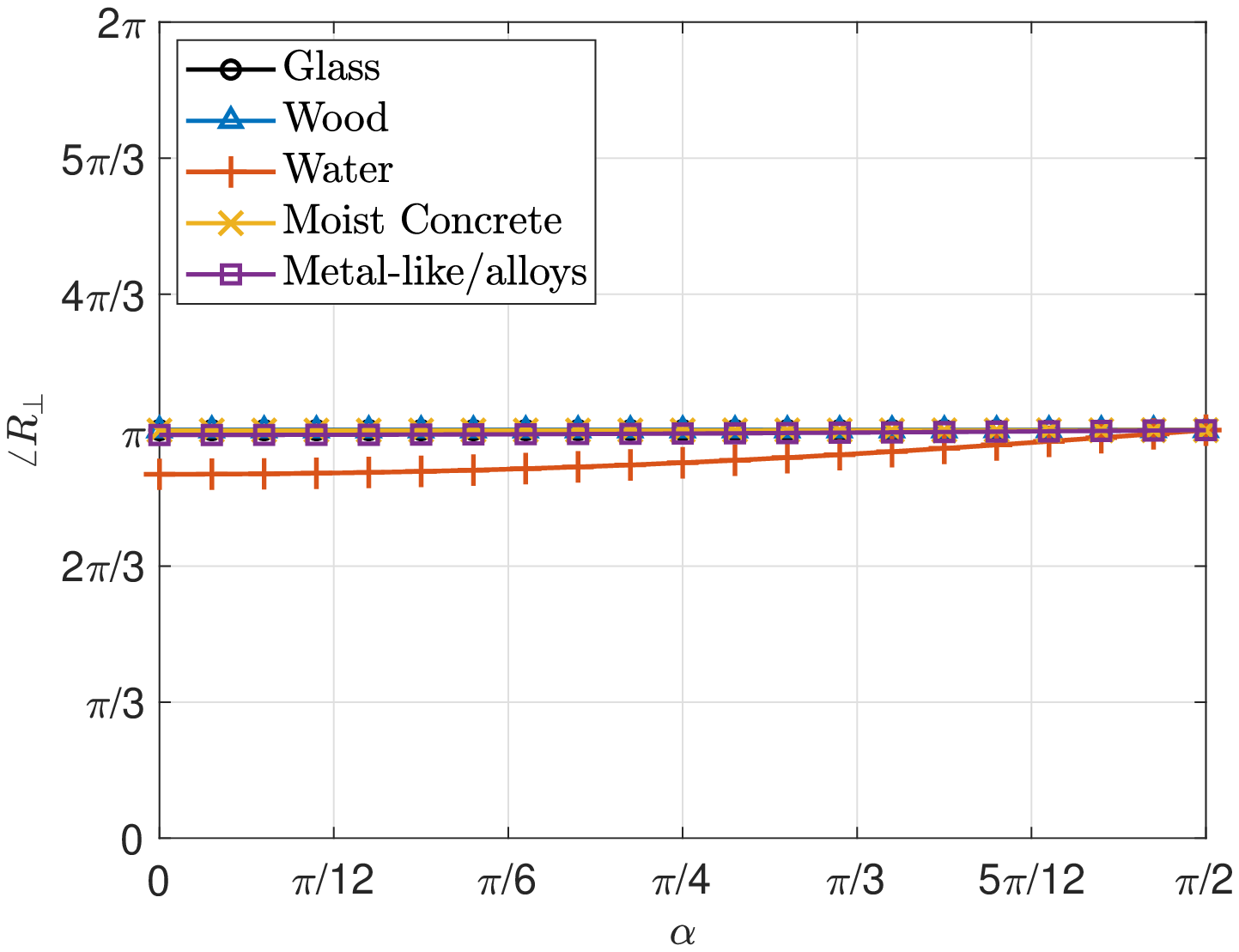}
	\caption{Perpendicular reflection coefficient of the materials in Table~\ref{table1} as a function of incidence angle $\alpha$ and specified by amplitude (top) and phase (bottom).}
	\label{Fig4}
\end{figure}
\subsection{Carrier Phase Measurements}\label{carrier}
Assume that the UE employs a phase-locked loop circuit to track the phase of the tone sent by the AP at frequency~$f$. Then, the phase measurements at the UE can be written as
\begin{align}
\varphi=\mathrm{mod}\Big(\frac{2\pi}{\lambda}d \ +\ \varphi^\circ + \tilde{\varphi} + \bar{\varphi} ,2\pi\Big),\label{phase} 
\end{align}
where $ \varphi^\circ$ is the difference between the AP and UE clock phase offsets, $\tilde{\varphi}$ is the phase measurement noise, and $\bar{\varphi}$ is the phase shift introduced by the polarization configuration and/or NLOS reflections. Moreover, recall that $d$ is the distance traveled by the tone. Therefore, $d$ matches either i) the Euclidean distance between the AP and UE in the case of LOS, or ii) the sum of the Euclidean distances between AP-reflector and reflector-UE in the case of NLOS.

The measurement noise is modeled as a zero mean normal random variable, i.e., $\tilde{\varphi} \sim \mathcal{N}(0,\sigma^2)$, with the variance given by \cite[eq.~(1.15)]{Henkel.2010}
\begin{align}
\sigma^2 = \frac{1}{\gamma}\Big(1+\frac{1}{TB\gamma}\Big).\label{sigma}
\end{align}
Here, $B$ is the bandwidth of the loop filter, and $T$ is the integration time. Meanwhile, $\gamma$ is the signal-to-noise ratio (SNR) at the receive antenna, and is given by \cite[Ch.2]{Balanis.2015}
\begin{align}
\gamma = \frac{G\lambda^2\big|E_r\big|^2}{480\pi^2BN_0},\label{gamma}
\end{align}  
where $N_0$ is the noise spectral density, and $G$ is the UE's antenna gain. 
Finally, $\bar{\varphi}$ is given by
\begin{align}
\bar{\varphi} =  \text{arctan2}(\Im\{\mathbf{F}_r^T\mathbf{M}\mathbf{F}_t\},\Re\{\mathbf{F}_r^T\mathbf{M}\mathbf{F}_t\}).
\end{align}
\section{Hypothesis Testing Design}\label{hypT}
The experiment is designed as follows. AP's/UE's antennas are configured to transmit/receive with either horizontal ($H$) or vertical ($V$) polarization. Thus, there are $4$ or $2$ possible combinations in a full-diversity (where both transmit and receive antennas are dual-polarized) or transmit/receive (only)-diversity system, respectively. Let us use the super-index pair $(p,q)$, with $p,q\in\{H,V\}$, to indicate each possible  polarization configuration. In this case, $p$ and $q$ correspond to the receive and transmit polarization configuration, respectively. Now, we define matrix $\mathbf{E}_r$ to collect all the possible polarization configurations as
\begin{align}
\mathbf{E}_r=\begin{bmatrix}
E_r^{V,V} & E_r^{V,H} \\
E_r^{H,V} & E_r^{H,H} 
\end{bmatrix}.
\end{align}
Note that a pure vertical, horizontal configuration implies that
$\mathbf{F_u}=[F_u^V(\phi^u,\theta^u),0]^T$, $\mathbf{F_u}=[0,F_u^H(\phi^u,\theta^u)]^T$, respectively. Therefore, the rotation angle for each possible polarization configuration may be different ($\vartheta$ for LOS, and $\vartheta_1,\vartheta_2$ for NLOS, are different). Finally, carrier phase measurements are collected using a time-division protocol, i.e., measuring over one polarization configuration at a time.
\subsection{LOS/NLOS Hypothesis}
By substituting \eqref{M} into \eqref{Er} for each possible polarization configuration, one obtains 
\begin{align}
\mathbf{E}_r^\text{los}& = \zeta\begin{bmatrix}
\eta_1\cos\vartheta^{V,V}  & \eta_2\sin\vartheta^{V,H}  \\
-\eta_3\sin\vartheta^{H,V} & \eta_4\cos\vartheta^{H,H} &  \\
\end{bmatrix}\label{ErM}
\end{align}
under LOS propagation conditions, where
\begin{subequations}
\begin{alignat}{2}
\eta_1&\triangleq F_r^V(\phi^r,\theta^r)F_t^V(\phi^t,\theta^t),\label{eta4}\\
\eta_2&\triangleq F_r^V(\phi^r,\theta^r)F_t^H(\phi^t,\theta^t),\\
\eta_3&\triangleq F_r^H(\phi^r,\theta^r)F_t^V(\phi^t,\theta^t),\\
\eta_4&\triangleq F_r^H(\phi^r,\theta^r)F_t^H(\phi^t,\theta^t).\label{eta1}
\end{alignat}
\end{subequations}
Similar to~\eqref{ErM}, one can obtain $\mathbf{E}_r^\text{nlos}$ in \eqref{ErMnlos} (at the top of the next page) under NLOS propagation conditions by substituting \eqref{Mnlos} into \eqref{Er}. Finally, let the transmit/receive beamformers be rotated to offset the phases of $\mathbf{F}_u$, $\forall u\in\{r,t\}$, such that $\eta_1,\eta_2,\eta_3,\eta_4\in \mathbb{R}^+$.
\begin{figure*}[h]
\begin{align}
\mathbf{E}_r^\text{nlos}\!\! =\! \zeta\!\begin{bmatrix}
 \eta_1(-R_\parallel\sin\vartheta_1^{V,V}\!\sin\vartheta_2^{V,V}\!+\!R_\perp\cos\vartheta_1^{V,V}\!\cos\vartheta_2^{V,V})
&\! \eta_2(R_\parallel\cos\vartheta_1^{V,H}\!\sin\vartheta_2^{V,H}\!+\!R_\perp\sin\vartheta_1^{V,H}\!\cos\vartheta_2^{V,H})\\
-\eta_3(R_\parallel\sin\vartheta_1^{H,V}\!\cos\vartheta_2^{H,V}\!+\!R_\perp\cos\vartheta_1^{H,V}\!\sin\vartheta_2^{H,V}) &\! \eta_4(R_\parallel\cos\vartheta_1^{H,H}\!\cos\vartheta_2^{H,H}\!-\!R_\perp\sin\vartheta_1^{H,H}\!\sin\vartheta_2^{H,H})
\end{bmatrix}\!. \label{ErMnlos}
\end{align} \hrule
\end{figure*}
\subsection{Relevant Decision Metric}\label{relevantMetric}
We assume the UE's antenna array rotation configuration is unknown. Note that in scenarios where the UE is aware of the AP's antenna orientation and is equipped with proper sensing mechanisms, e.g., gyroscope and geomagnetic sensors, the value of $\beta_x,\beta_y,\beta_z$ may be (approximately) known beforehand and exploited for LOS/NLOS identification, which we leave for future work. In such scenarios, the imperfection in the estimation of these parameters, which may be significant if the UE is a low-capability device, must be considered.
 
In order to remove the impact of the clock phase offsets $\varphi^\circ$ and the phase shift introduced by the path distance $d$, which are unknown but common to all the phase measurements, we propose the following relevant decision metric based on differential measurements:
\begin{align}
\bm{\vartriangle}_{i}=&\mathrm{mod}\Big([
\varphi^{V,V} \!\!\!-\! \varphi^{V,H}\!\!,
\varphi^{V,V} \!\!\!-\! \varphi^{H,V}\!\!,
\varphi^{V,V} \!\!\!-\! \varphi^{H,H}\!\!, 
\varphi^{V,H} \!\!\!-\! \varphi^{H,V}\!\!,\nonumber\\
&\qquad\ \ \ 
\varphi^{V,H} \!\!\!-\! \varphi^{H,H}\!\!,
\varphi^{H,V} \!\!\!-\! \varphi^{H,H}
]^T
\!\!+
\!\pi\bm{\nu}_{i}, 2\pi\Big).\label{Lambda} 
\end{align}
Observe that $\bm{\vartriangle}_{i}$ in \eqref{Lambda} exploits the measurements over all the  four polarization configurations. Here, $\bm{\nu}_i\in[0,1]^{6\times 1}$  is a binary 6-element vector representation of the decimal number $i$, thus,  $i\in\{0,1,2,\cdots,2^6-1\}$. For instance, $\bm{\nu}_4=[0, 0, 0, 1, 0, 0]^T$, while $\bm{\nu}_{19}=[0, 1, 0, 0, 1, 1]^T$. 
The role of $\pi\bm{\nu}_{i}$ in \eqref{Lambda} is explained next.

Observe that $\mathbf{E}_r^\text{los}/\zeta\in\mathbb{R}^{2\times 2}$, while $\mathbf{E}_r^\text{nlos}/\zeta\in\mathbb{C}^{2\times 2}$ due to complex valued reflection coefficients \eqref{Rper}, \eqref{Rpar}. This means that the set of phase shifts that could be  introduced under NLOS propagation conditions is  infinite, i.e., $\bar{\varphi}\in[0,2\pi]$, while they are limited to $\bar{\varphi}\in\{0,\pi\}$ under LOS. Then, under LOS conditions, there is a vector 
$\pi\bm{\nu}_i=[0,\pi]^{6\times 1}$ that ideally drives \eqref{Lambda} to zero.
In other words, under LOS conditions and perfect measurements, one obtains $\bm{\vartriangle}_i=\mathbf{0}$ for a certain $i\in\{0,1,2,\cdots,63\}$ according to the specific UE's antenna rotation configuration. Meanwhile, under NLOS conditions, $\bm{\vartriangle}_{i}$ depends on the specific reflector orientation and reflection coefficients, thus, will be surely different from $\mathbf{0}$ independently of the UE's antenna array rotation configuration. In all the cases, the common (path) phase shift $2\pi d/\pi$ and the clock phase offsets $\varphi^\circ$ are removed by the differential operation.

Based on the above formulation, the hypothesis testing can be designed such that
$\min_{i}\lVert\bm{\vartriangle}_i\rVert^2=0$ implies LOS ($H_0$), while $\min_{i}\lVert\bm{\vartriangle}_i\rVert^2> 0$ indicates NLOS ($H_1$). However, measurements are imperfect in practice due to noise and other non-modeled phenomena such as scattering. Therefore, the hypothesis needs to be relaxed as
\begin{align}
\left\{\begin{array}{ll}
\min_{i}\lVert\mathbf{W}\hat{\bm{\vartriangle}}_i\rVert^2\le \xi,&\rightarrow \text{declare LOS } (H_0) \\
\min_{i}\lVert\mathbf{W}\hat{\bm{\vartriangle}}_i\rVert^2> \xi,&\rightarrow \text{declare NLOS } (H_1) 
\end{array}\right.\!\!\!,\label{dec}
\end{align}
where $\hat{\bm{\vartriangle}}_i$ denotes the estimation of $\bm{\vartriangle}_i$ based on measurements, and $\xi$ is a decision threshold. The greater the threshold value is, the greater the LOS detection probability is, but also false-alarm probabilities become more prominent.
Therefore, $\xi$ must be carefully selected for enabling accurate detection capabilities. Unfortunately, analytically obtaining the optimum threshold inevitably requires deriving the statistics of $\min_{i}\lVert\mathbf{W}\hat{\bm{\vartriangle}}_i\rVert^2$ under LOS and NLOS conditions, which is a cumbersome task, hence, we resort to numerical methods in Section~\ref{results}. Nevertheless, one can rely on simulation campaigns, numerical optimization methods, and ML techniques to (sub-optimally) set $\xi$ in practice. Meanwhile,
$\mathbf{W}$ is a diagonal real weighting matrix with entries $w_j, j=1,\cdots,6$ and $\lVert\mathbf{W}\rVert_F^2=\sum_{j=1}^{6}w_j^2=1$ enabling a differential treatment of the elements of $\hat{\bm{\vartriangle}}_i$.
Observe that by zeroing a certain $w_j$, one can null the effect of any of the elements of $\hat{\bm{\vartriangle}}_{i}$ (and $\bm{\vartriangle}_i$). This can be exploited for assessing the performance of systems with either transmit or receive polarization diversity. We provide some numerical discussions on this in Section~\ref{results}.

A downside of the above approach is that $\Im\{\mathbf{E}_r^\text{los}/\zeta\}, \Re\{\mathbf{E}_r^\text{los}/\zeta\} \approx 0$ for a wide range of reflection-related parameters. This is because $60\kappa\lambda$ in $\varepsilon$ (check \eqref{Rper} and \eqref{Rpar}) is very small at the mm-wave spectrum operation (i.e., small $\lambda$) and/or when reflectors are dielectric (i.e., small $\kappa$). In such scenarios, the angle of $\mathbf{E}_r^\text{los}$ is very close to 0 or $\pi$, which affects the performance of a detector based on \eqref{Lambda}--\eqref{dec}. We address this issue as discussed in the following.

Note that not every $\bm{\nu}_i,\ i=0, \cdots, 63$, needs to be evaluated in LOS scenarios due to some geometric relations that can be exploited. Specifically, $\vartheta^{V,V}\!=\vartheta^{V,H}$ and $\vartheta^{H,V}\!=\vartheta^{H,H}$ holds under LOS, while there are even some spatial geometric relationships between $\vartheta^{V,V}$ and $\vartheta^{H,H}$ which are difficult to derive in closed-form, but can be obtained numerically using the framework in Section~\ref{LOS}. In a nutshell, there are some vectors $\bm{\nu}_i$ that would never make $\bm{\vartriangle}_i=\bm{0}$ when operating in LOS conditions. In fact, it can be shown that out of the $64$ possible vectors $\bm{\nu}_i$, one (and only one) of the $8$ following vectors: $\bm{\nu}_{0}, \bm{\nu}_{11}, \bm{\nu}_{21}, \bm{\nu}_{30}, \bm{\nu}_{38},\ \bm{\nu}_{45},\ \bm{\nu}_{51},\ \bm{\nu}_{56}$,  makes $\bm{\vartriangle}_i=\bm{0}$ in LOS conditions with perfect phase measurements. Therefore, we use only these vectors when evaluating \eqref{Lambda}.
\subsection{On the Configuration of the Weights}\label{weightsC}
By substituting \eqref{phase} into \eqref{Lambda}, one obtains 
\begin{align}
	\hat{\bm{\vartriangle}}_{i}\!=\! \mathrm{mod}\!\left(\!\begin{bmatrix}
	\tilde{\varphi}^{V,V}\! +\! \bar{\varphi}^{V,V}\! -\! \tilde{\varphi}^{V,H}\! -\! \bar{\varphi}^{V,H}\\
		\tilde{\varphi}^{V,V} \!+\! \bar{\varphi}^{V,V} \!-\! \tilde{\varphi}^{H,V}
		-\! \bar{\varphi}^{H,V} \\
	\tilde{\varphi}^{V,V} \!+\! \bar{\varphi}^{V,V} \!-\! \tilde{\varphi}^{H,H} \!-\! \bar{\varphi}^{H,H}\\
	\tilde{\varphi}^{V,H} \!+\! \bar{\varphi}^{V,H}
	- \tilde{\varphi}^{H,V}\!-\! \bar{\varphi}^{H,V}\\
	\tilde{\varphi}^{V,H} \!+\! \bar{\varphi}^{V,H}\!-\! \tilde{\varphi}^{H,H} \!-\! \bar{\varphi}^{H,H}\\
	\tilde{\varphi}^{H,V} \!+\! \bar{\varphi}^{H,V} \!-\! \tilde{\varphi}^{H,H}\!-\! \bar{\varphi}^{H,H}
	\end{bmatrix}
	\!\!\!+\!\pi\bm{\nu}_{i}, 2\pi\!\right)\!. \label{That}  
\end{align}
%
Note that due to the intricacy of \eqref{ErMnlos}, and the several unknowns, i.e., $R_\parallel$, $R_\perp$, and reflector orientation, it becomes cumbersome getting specific insights for the attainable values of $\min_{i}\lVert\mathbf{W}\hat{\bm{\vartriangle}}_i\rVert^2$ for the NLOS scenario. Therefore, we focus merely on the LOS scenario in the following.

Under LOS conditions, we have that
\begin{align}
    &\big[\bar{\varphi}^{V,V}-\bar{\varphi}^{V,H},\bar{\varphi}^{V,V}-\bar{\varphi}^{H,V},\bar{\varphi}^{V,V}-\bar{\varphi}^{H,H},\bar{\varphi}^{V,H}-\bar{\varphi}^{H,V},\nonumber\\
    &\qquad\qquad \bar{\varphi}^{V,H}-\bar{\varphi}^{H,H},\bar{\varphi}^{H,V}-\bar{\varphi}^{H,H}\big]^T+\pi\bm{\nu}_i=\bm{0}\label{eqP}
\end{align}
for a certain $i\in\{0,11,21,30,38,45,51,56\}$. Then, by using \eqref{eqP} when evaluating \eqref{That},
and expanding the squared norm, one obtains 
\begin{align}
\min_{i} \lVert\mathbf{W}\hat{\bm{\vartriangle}}_{i}\rVert^2&\!= w_1^2\big(\tilde{\varphi}^{V,V} \!\!\! -\!  \tilde{\varphi}^{V,H}\big)^2\!\!\! +\!
w_2^2\big(\tilde{\varphi}^{V,V}\!\!\!  -\!  \tilde{\varphi}^{H,V}\big)^2\nonumber\\
&+\!w_3^2\big(\tilde{\varphi}^{V,V} \!\!\! -\! \tilde{\varphi}^{H,H}\big)^2
\!\!+\!w_4^2\big(\tilde{\varphi}^{V,H}\!\!\!-\!  \tilde{\varphi}^{H,V}\big)^2\nonumber\\
&+\!w_5^2\big(\tilde{\varphi}^{V,H}\!\!\!-\!  \tilde{\varphi}^{H,H}\big)^2\!\!\!+\!w_6^2\big(\tilde{\varphi}^{H,V}\!\!\!-\!  \tilde{\varphi}^{H,H}\big)^2\label{Tlos}
\end{align}
 assuming LOS and relatively accurate phase measurements. In the following, we discuss three different weighting approaches. Notice that we cannot claim any optimality for these weighting approaches as this would require exploiting the statistics of  $\min_{i} \lVert\mathbf{W}\hat{\bm{\vartriangle}}_{i}\rVert^2$ in both LOS and NLOS scenarios.
\subsubsection{Equal (EQU) weighting}
This is the simplest approach. Each addend element of $\hat{\bm{\vartriangle}}_{i}$ is equally weighted, thus,
\begin{align}
    w_j=1/\sqrt{6},\ \forall j. \label{weq}
\end{align}
\subsubsection{Minimum noise variance (MNV) weighting}
Intuitively, a proper weighting configuration should take into account the measurements accuracy. 
Assume that the variance of the noise measurements is known. Indeed, such statistics can be readily estimated based on SNR measurements followed by the evaluation of \eqref{sigma}. 
Now, observe that by departing from \eqref{Tlos}, one can write
\begin{align}
  \mathbb{E}\Big[\!\min_{i} \lVert\mathbf{W}\hat{\bm{\vartriangle}}_{i}\rVert^2\!\Big] \!\!=\! \sum_{j=1}^6\! \sigma^2_j w_j^2 \!\ge\! \min_z \sigma^2_z \sum_{j=1}^6\! w_j^2\!=\!\min_z \sigma^2_z, \label{sumsq}
\end{align}
where 
\begin{subequations}
\begin{alignat}{2}
    \sigma^2_1&\triangleq (\sigma^{V,V})^2+(\sigma^{V,H})^2,\\
    \sigma^2_2&\triangleq (\sigma^{V,V})^2+(\sigma^{H,V})^2,\\
    \sigma^2_3&\triangleq (\sigma^{V,V})^2+(\sigma^{H,H})^2,\\
    \sigma^2_4&\triangleq (\sigma^{V,H})^2+(\sigma^{H,V})^2,\\
    \sigma^2_5&\triangleq (\sigma^{V,H})^2+(\sigma^{H,H})^2,\\
    \sigma^2_6&\triangleq (\sigma^{H,V})^2+(\sigma^{H,H})^2,
\end{alignat}\label{sigEq}
\end{subequations}
\!while the last equality in \eqref{sumsq} comes from using $\sum_{j=1}^{6}w_j^2\!=\!1$. Notice that the lower bound in \eqref{sumsq}, i.e., the MNV weighting configuration, is obtained from setting
\begin{align}
    w_j^\star=\left\{
    \begin{array}{cl}
         1,& \text{for } j=\arg\min_z \sigma^2_z   \\
         0,& \text{otherwise}. 
    \end{array}
    \right. \label{wjopt}
\end{align}
This is, only the term of \eqref{Tlos} with the best expected accuracy is considered for the LOS/NLOS detection.
\subsubsection{Noise variance -proportional (NVP) weighting}\label{nvpV}
The main problem with the MNV approach is that
it seriously constrains the exploitation of $\bm{\nu}_i$ for LOS/NLOS detection. Note that five out of the six elements of $\bm{\nu}_i$ are neglected when using the MNV weighting, thus, one cannot take full advantage of its particular construction structure for noise mitigation. Herein, we circumvent this issue by adopting squared weights inversely proportional to the noise variances. Specifically,
\begin{align}
w_j^\star&=\mu/\sigma_j,\label{wj}
\end{align}
where $\sigma_j$ can be obtained from \eqref{sigEq} for $j\in\{1,2,3,4,5,6\}$, and $\mu$ is a normalization factor that guarantees 
$\sum_{j=1}^6\!w_j^2\!=\!1$,
thus, it is given by
\begin{align}
\mu &= \frac{1}{\sqrt{\sum_{j=1}^6\sigma_j^{-2}}}.\label{mu}
\end{align}
In this way, the noisier elements of $\hat{\bm{\vartriangle}}_{i}$
weigh less, but they are never neglected unless operating in the high SNR asymptotic regime. Nevertheless, it should be noted that compared to the MNV weighting  in \eqref{wjopt}, the performance under the NVP weighting in \eqref{wj} is more prone to degradation due to imperfect estimations of the measurement noise variances. That is because configuring the MNV weights requires only a relative, potentially rough, estimate of the measurement noise variances $\{\sigma_j^2\}$ (specifically, determining $\arg\min_j \sigma_j^2$), while the specific measurement noise variances matter for computing the NVP weights.

The LOS/NLOS detection performance under the above weighting approaches is numerically assessed in the following.
\section{Numerical Results}\label{results}
Here, we numerically illustrate the performance of the proposed LOS/NLOS identification method. 
We produce $10^5$ scenario instances comprising LOS and NLOS configurations with equal probability. We generate $\alpha\in[0,\pi/2]$, $\beta_x,\beta_y,\beta_z,\delta_x,\delta_y,\delta_z\in[-\pi,\pi]$,  $\phi^t, \phi^r\in[-\pi,\pi]$, and $\theta^t, \theta^r\in[-\pi/2,\pi/2]$  uniformly random, and set $B=20$ Hz, $T=10$ ms,  $N_0=10^{-20.38}$~W/Hz assuming ambient temperature of 300~K,  $G=0$ dB, $f=30$ GHz.\footnote{Notice that the proposed LOS/NLOS classifier does not strictly rely on any prior information about these and other parameters (e.g., $P$, and $\bar{\bar{\sigma}}^2$ in Fig.~\ref{Fig7}) as it is purely based on measurements. However, a prior info on the noise power, i.e., $BN_0$, may be beneficial for estimating the measurement noise variances via \eqref{sigma}, thus, facilitating the implementation of the NVP weighting approach. In any case, information about the noise power is usually available/measurable in practice.} 
Finally, we assume $F_u^H(\phi^u,\theta^u)=1$ and $F_u^V(\phi^u,\theta^u)=1$ for horizontally and vertically polarized transmissions ($u=t$) or receptions ($u=r$) such that the spatial antenna gains/losses (antenna pattern) are ignored or counteracted by beamforming gains.
\subsection{On the Probabilities of Miss Detection and False Alarm}\label{probMiss}
Herein, we assess the performance of the proposed detection mechanism by evaluating the probability of miss-detection (PMD) and probability of false alarm (PFA). Specifically, the LOS -- PMD and PFA are respectively given by 
\begin{itemize}
	\item the probability of declaring NLOS given LOS conditions: $\Pr(H_1|H_0)$;
	\item the probability of declaring LOS given NLOS conditions: $\Pr(H_0|H_1)$.
\end{itemize}
Observe that the LOS -- PMD and PFA match respectively the NLOS -- PFA and PMD, thus, we merely focus on the former pair in the following. 

\begin{figure}[t!]
	\centering
	\includegraphics[width=0.46\textwidth]{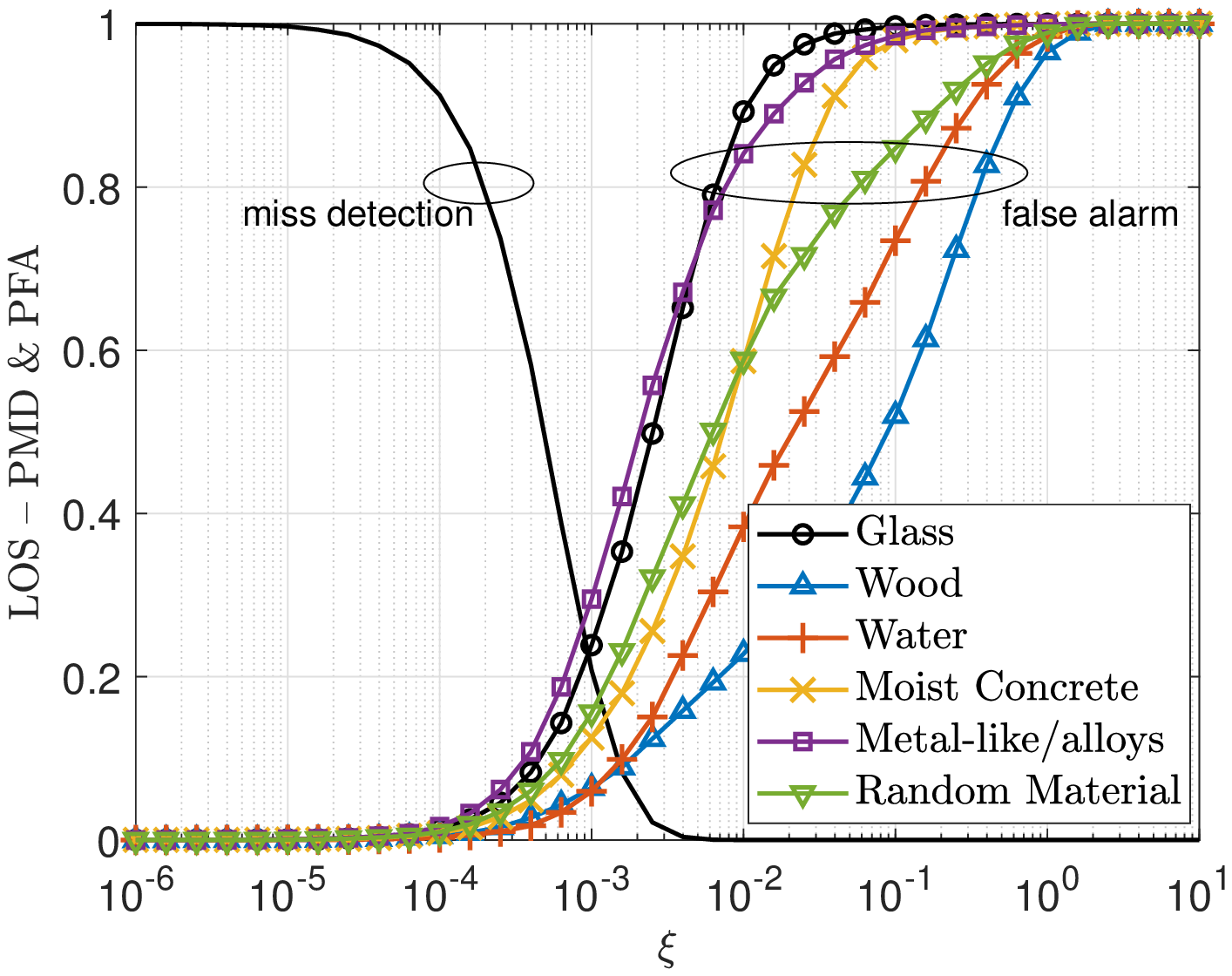}\\
	\includegraphics[width=0.46\textwidth]{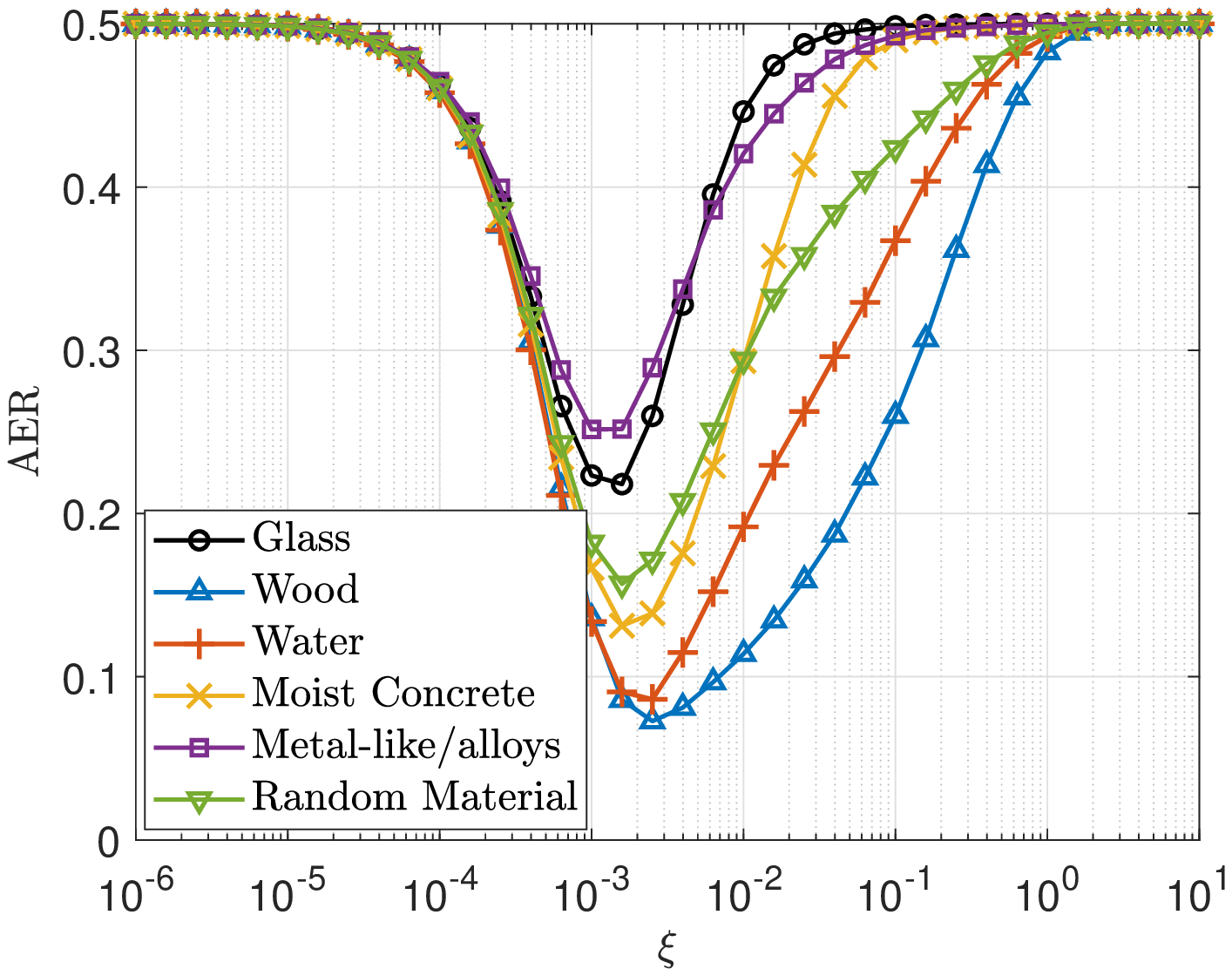}
	\caption{a) LOS -- PMD and PFA (top), and b) AER (bottom) as a function of $\xi$. We use the NVP weighting approach, and set
	$P=-100$~dB.}
	\label{Fig5}
\end{figure}

Fig.~\ref{Fig5} evaluates the performance of a full-diversity system (both transmit and receive polarization diversity) as a function of the decision threshold $\xi$. Herein, we adopt the NVP
weighting approach, while the EQU and MNV weighting configurations and their relative performance are analyzed later in Sections~\ref{weightsConf} and \ref{Scatt}.
Moreover, in the case of NLOS, we assume that a reflector is composed of one of the materials given in Table~\ref{table1}. Notice that besides the performance for specific reflecting materials, we also show the performance for a randomly chosen (among those listed in Table~\ref{table1}) material.
 
Fig.~\ref{Fig5}a shows the LOS -- PMD and PFA performance curves, which are in agreement with our arguments in Section~\ref{relevantMetric}: a relatively small (large) $\xi$ leads to a small LOS--PFA (LOS--PMD) but a large LOS--PMD (LOS--PFA). Meanwhile, Fig.~\ref{Fig5}b illustrates the AER performance metric, which is given by
\begin{align}
\text{AER} &= \Pr(H_1|H_0)\Pr(H_0) + \Pr(H_0|H_1)\Pr(H_1) \nonumber\\
&= 0.5\times(\Pr(H_1|H_0)+\Pr(H_0|H_1)),\label{AER}
\end{align}
where we exploit $\Pr(H_0)=\Pr(H_1)=0.5$ in the last step. As expected and shown in Fig.~\ref{Fig5}b, the optimum performance in terms of AER is obtained when both the LOS--PFA and LOS--PMD are relatively favorable.

Observe that there is a unique curve for the LOS -- PMD in Fig.~\ref{Fig5}a since there are no reflectors in LOS scenarios. Meanwhile, the LOS -- PFA is different for each reflecting material as it depends on its corresponding electric characteristics $\{\epsilon_r,\kappa\}$, which influence differently the reflection coefficients $\{R_\perp,R_\parallel\}$ as illustrated in Figs.~\ref{Fig3} and \ref{Fig4}. The detection capabilities in an environment with reflections from water and/or wooden surfaces are superior than in the other considered environments, specially compared to the environment with metallic reflectors. To understand this behavior, please refer to Figs.~\ref{Fig3} and \ref{Fig4}, and note that the wooden reflectors introduce considerable power losses to both polarization components of the incident signal, thus, promoting noisier measurements in NLOS conditions. Meanwhile, in the case of water reflection, the performance gains come from the continuous range of phase shifts (instead of approximately discrete as for other materials) that are introduced during the reflection of both vertically and horizontally polarized signal components. As corroborated in Fig.~\ref{Fig5}b, the relative performance in terms of AER is determined by the LOS -- PFA. Interestingly, the optimum decision threshold remains around the same value for all the considered reflecting materials, i.e., $10^{-3}\le \xi^\text{opt}\le 3\times 10^{-3}$, although being slightly larger for reflecting materials promoting better performance, e.g., wood and water. 
\subsection{On the Performance Impact of the Path Power Gain, Limited-Diversity, and Weighting Configuration}\label{weightsConf}
In the following, we focus only on the AER metric. Specifically, we illustrate the performance of the optimum AER, which is the minimum  achievable AER for any $\xi$. As commented in Section~\ref{relevantMetric}, analytically obtaining the optimal $\xi$ is a cumbersome task, hence we resort to a numeric brute-force optimization here, but more evolved numerical optimization methods, simulation campaigns, and ML techniques can be designed to (sub-optimally) set $\xi$ in practice.
\begin{figure}[t!]
	\centering
	\ \ \ \ \ \ \ \includegraphics[width=0.46\textwidth]{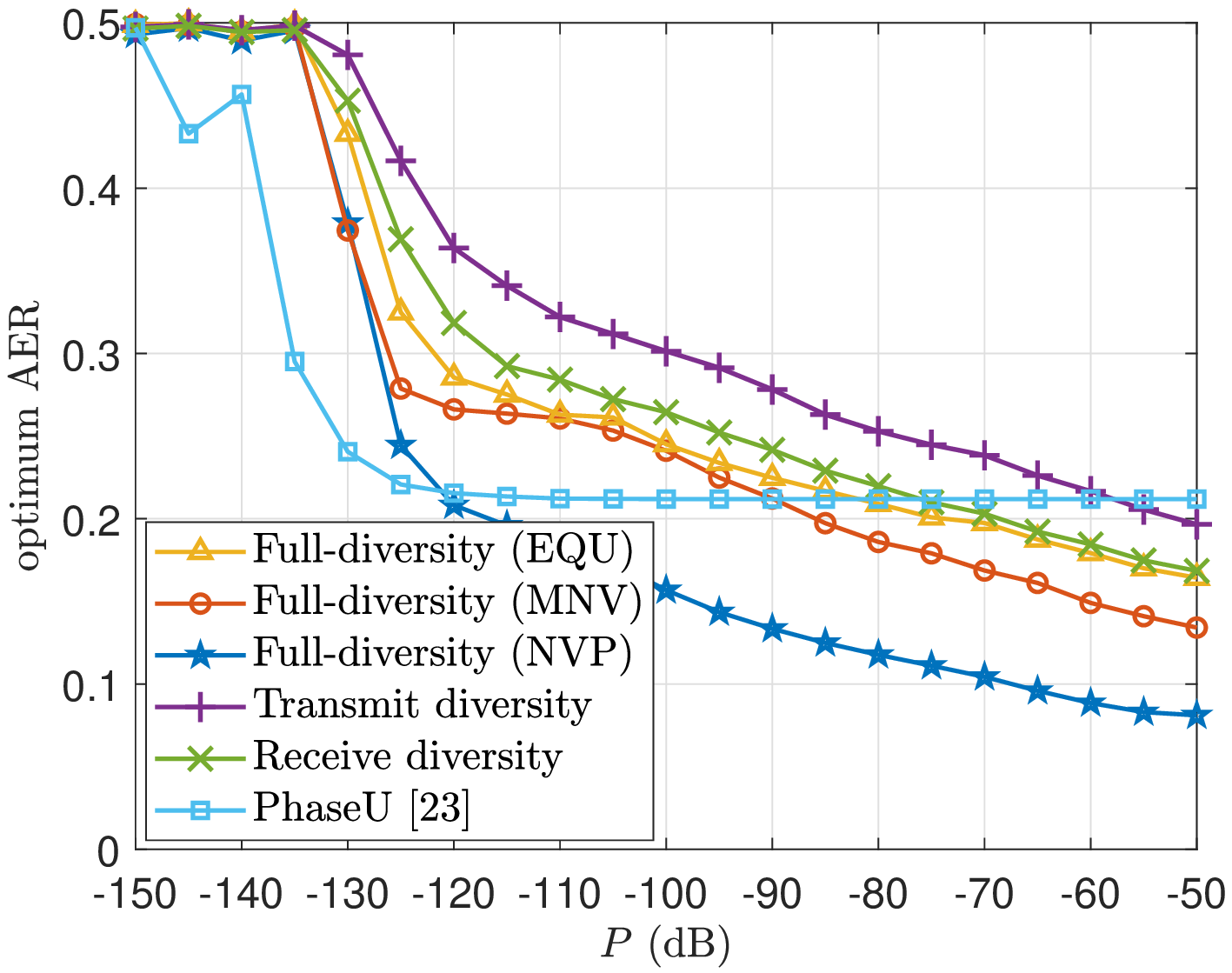}\\
	\!\!\!\!\!\!\!\includegraphics[width=0.47\textwidth]{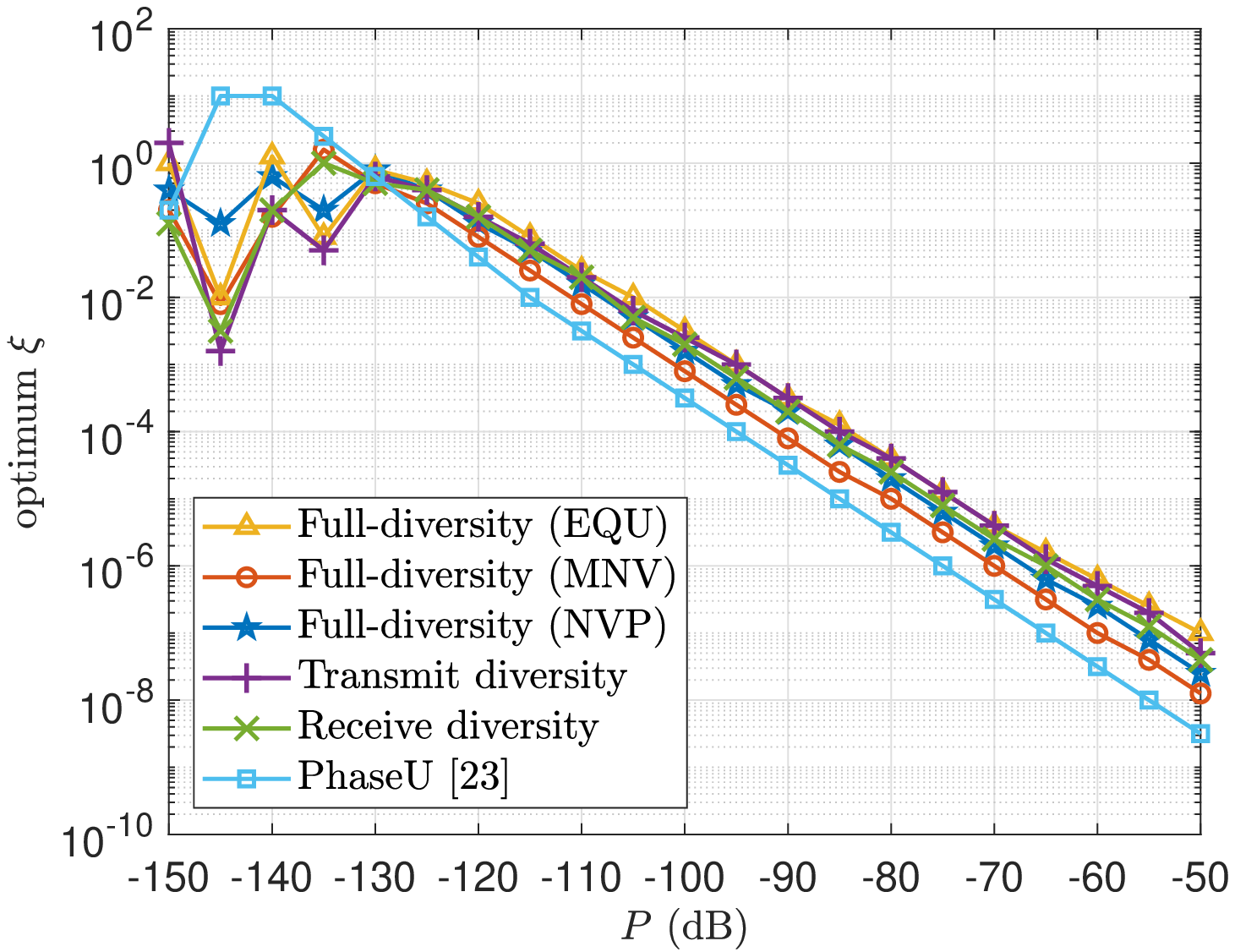}
	\caption{a) Optimum AER (top), and b) corresponding $\xi$ (bottom) as a function of the path power gain. In NLOS scenarios, the reflecting material is selected randomly among those given in Table~\ref{table1}.}
	\label{Fig6}
\end{figure}

We evaluate the performance of a limited-diversity system, where one of the link sides is assumed equipped with just a horizontally-polarized antenna. In such a case, the only measurements that can be collected are i)  $\varphi^{H,H}$ and $\varphi^{H,V}$ in the case of transmit diversity, and ii) $\varphi^{H,H}$ and $\varphi^{V,H}$ in the case of receive diversity. Therefore, only the last and penultimate elements of $\bm{\vartriangle}_{i}$ in \eqref{Lambda} can be acquired/measured when using transmit and receive diversity, respectively. As commented in Section~\ref{relevantMetric}, this can be artificially modeled by properly adjusting the weighting configuration, specifically
	\begin{itemize}
		\item $\mathbf{W}=\mathrm{diag}([0,0,0,0,0,1])$ for transmit diversity, 
		\item $\mathbf{W}=\mathrm{diag}([0,0,0,0,1,0])$ for receive diversity.
	\end{itemize}
Moreover, we simulate the performance of PhaseU, proposed in \cite{Chenshu.2015} for LOS/NLOS identification in WiFi networks, for benchmarking. In a nutshell, PhaseU relies on a weighted sum of the variance of the phase measurements for LOS/NLOS classification. For simplicity, we assume that the measurements' variance is perfectly known, and adopt the optimum decision threshold for comparison fairness. Finally, we select a random material among those given in Table~\ref{table1} for each NLOS scenario instance.
\begin{figure}[t!]
	\centering
	\includegraphics[width=0.47\textwidth]{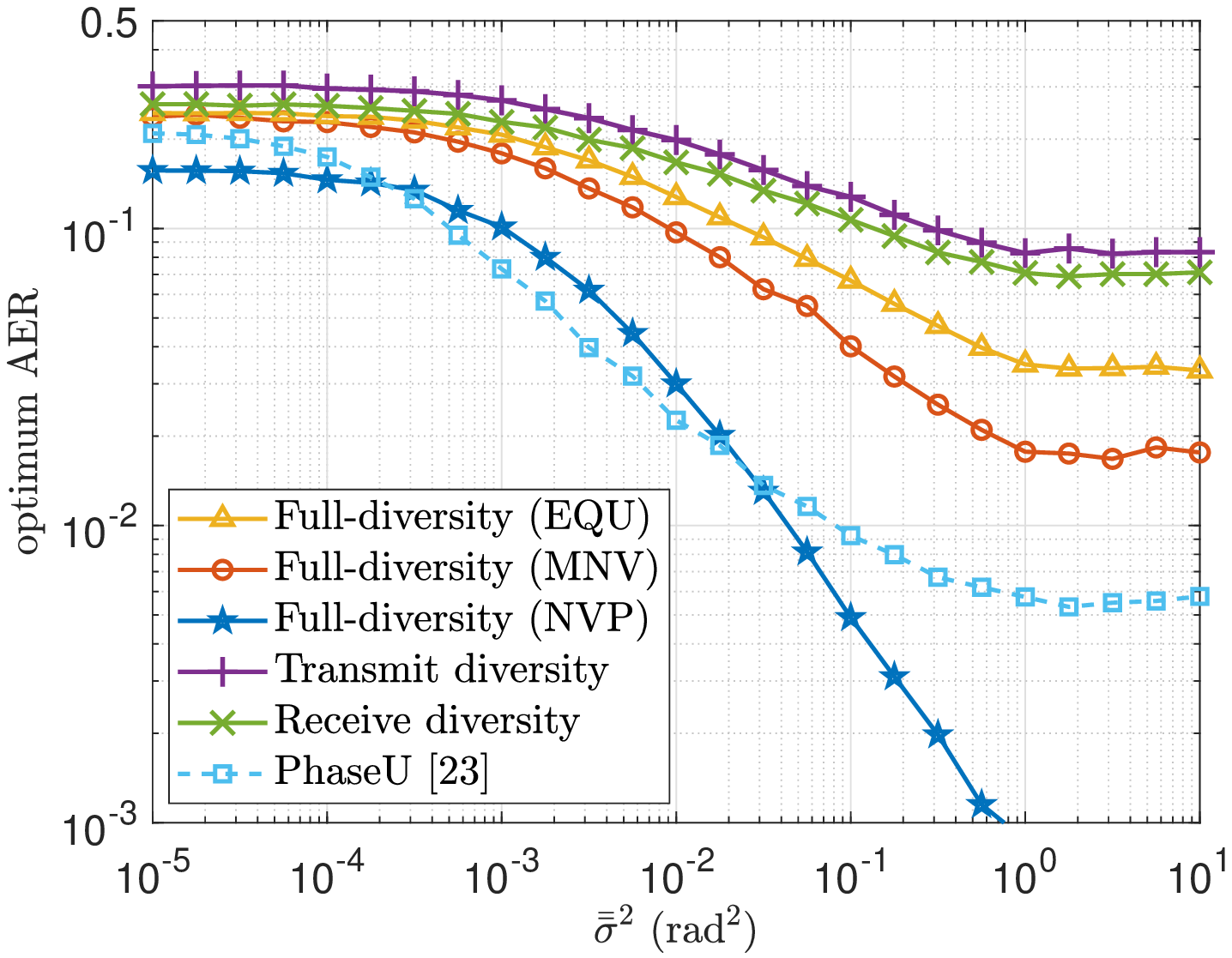}\\
	\includegraphics[width=0.47\textwidth]{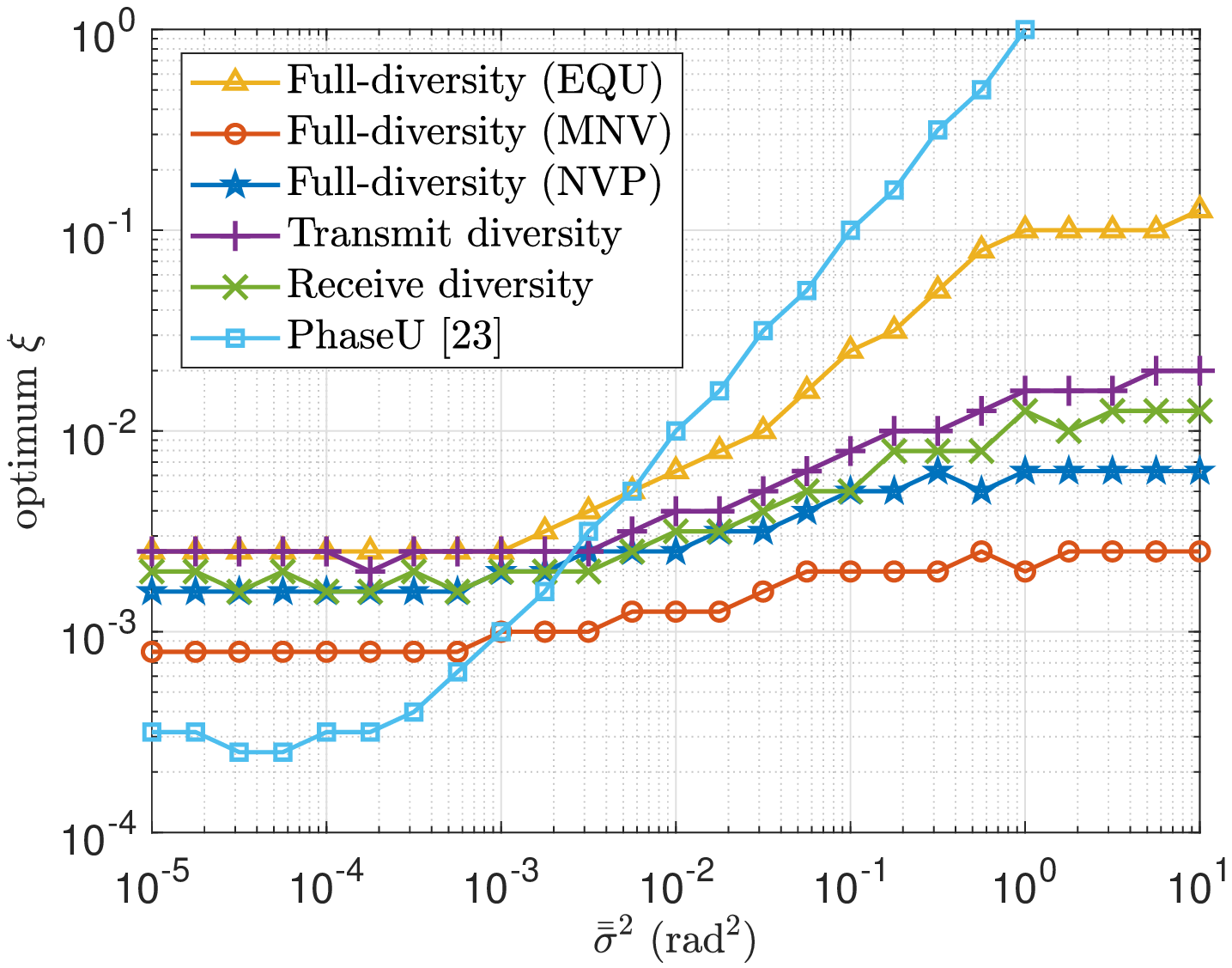}
	\caption{a) Optimum AER (top), and b) corresponding $\xi$ (bottom) as a function of the scattering phase noise power. We set $P=-100$~dB. The reflecting material in NLOS scenarios is selected randomly among those given in Table~\ref{table1}.}
	\label{Fig7}
\end{figure}

Fig.~\ref{Fig6} illustrates the system performance as a function of the path power gain $P$. Observe that the AER performance of the proposed approaches increases with $P$ since the SNR  of the received tones \eqref{gamma} is directly proportional to $|E_r|^2$, and thus to $P$,  consequently making phase measurements more accurate, i.e., with lower measurement noise variance according to \eqref{sigma}. However, the improvement rate is relatively slow for $P\ge-120$~dB  as shown in Fig.~\ref{Fig6}a, which evinces that the accuracy in this region is not mainly limited by the measurement noise but by potential similarities of the LOS and NLOS propagation in certain scenarios. Indeed, the AER is approximately decreased by  0.1 when the path power gain increases from $-120$ dB to $-70$ dB. Meanwhile, the AER performance of PhaseU completely saturates around $0.21$ for $P\ge-120$~dB, although it can still outperform the proposed approaches in scenarios with relatively low path power gain, specially given limited diversity and/or an unfavorable weighting configuration. Regarding the proposed weighting approaches, note that a full-diversity system with EQU  weighting does not provide significant performance improvements compared to a limited diversity system. Remarkably, significant performance gains are attainable by exploiting the statistics of the noise measurements, especially via NVP weighting. Indeed, a full-diversity system with NVP weighting  can significantly outperform all the other proposed configurations, including the one with MNV weighting. Meanwhile, the receive diversity outperforms the transmit diversity system, which may influence the link orientation design for LOS/NLOS identification in limited-diversity systems. Related to the decision threshold, Fig.~\ref{Fig6}b evinces that the optimum $\xi$, i.e., the one that minimizes the AER, remains approximately the same independently of the diversity type and/or weighting configuration for $P>-130$~dB, which is the operation region of interest. Note that the AER performance is extremely poor, close to 0.5, for $P<-130$~dB and independently of the $\xi$ choice, which is the reason for the divergence of the optimum $\xi$  in such operation region. Interestingly, the optimum $\xi$ decreases following a power-law as $P$ increases.
\subsection{On the Performance Impact of NLOS Scattering Noise and Number of Tones Transmitted}\label{Scatt}
Throughout the paper, we have ignored the multi-path interference (scattering) phenomenon  by assuming geometric mm-wave channels. In practice, NLOS scattering may not be completely ignored, thus, herein, we illustrate its potential performance impact. Specifically, we consider a scattering phase noise $\bar{\bar{\varphi}}\sim\mathcal{N}(0,\bar{\bar{\sigma}}^2)$ to be included as an additional addend to the right-side term in \eqref{phase} in NLOS conditions. 

\begin{figure}[t!]
	\centering
	\includegraphics[width=0.46\textwidth]{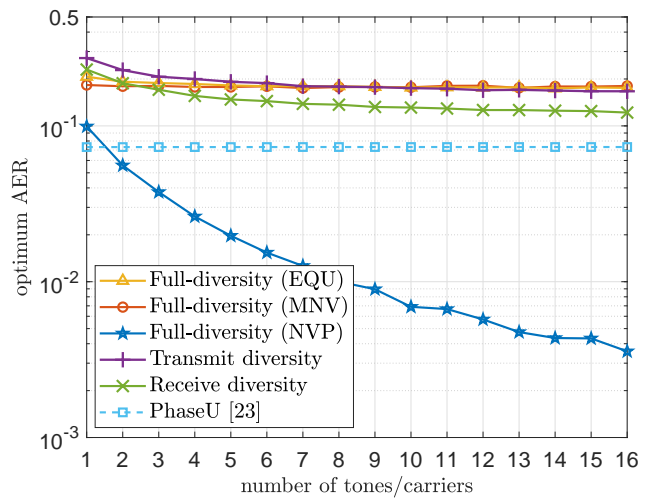}
	\caption{Optimum AER as a function of the number of tones transmitted. We set $P=-100$~dB and $\bar{\bar{\sigma}}^2=10^{-3}\ \mathrm{rad}^2$. Tones are spaced $1$ MHz, and the first tone is transmitted at $30$ GHz. In NLOS scenarios, the reflecting material is selected randomly among those given in Table~\ref{table1}.}
	\label{Fig8}
\end{figure}
Fig.~\ref{Fig7} illustrates the achievable optimum AER performance (Fig.~\ref{Fig7}a) and corresponding $\xi$ (Fig.~\ref{Fig7}b) as a function of $\bar{\bar{\sigma}}^2$.
 Observe that as $\bar{\bar{\sigma}}^2$ increases, the optimum AER decreases, while its corresponding $\xi$ increases since the NLOS hypothesis becomes noisier, thus, diverges more from the purely-geometric LOS hypothesis. The trends related to the relative performance  among full-diversity with equal/optimized weights and limited-diversity schemes remain similar to those observed in Fig.~\ref{Fig6}. It is also shown that PhaseU \cite{Chenshu.2015} always outperform the limited-diversity, and full-diversity with EQU and MNV weighting when $P=-100$ dB. Meanwhile, PhaseU only outperforms the proposed approach with full-diversity and NVP weighting for a limited set of $\bar{\bar{\sigma}}^2$ values, i.e., $\bar{\bar{\sigma}}^2\in[3\times 10^{-4}, 2\times 10^{-2}]\ \mathrm{rad}^2$ $\bar{\bar{\sigma}}^2$.

Throughout the paper, we have also assumed the transmission of a single tone/carrier, however, our framework can be easily extended to the multi-tone/carrier scenario. Specifically, one just needs to stack %
the relevant decision metric for each tone/carrier transmission \eqref{Lambda} to form an enlarged relevant decision metric $\bm{\vartriangle}_{i}$ of dimensions $6F\times 1$ for each $i$, where $F$ is the number of tones/carriers that are transmitted for carrier phase measurements. Moreover, the diagonal weighting matrix $\mathbf{W}$ has now $6F\times 6F$ elements, and can still be similarly designed  as described in Section~\ref{weightsC} according to an EQU, MNV or NVP approach.

Fig.~\ref{Fig8} shows the optimum AER performance as a function of the number $F$ of  tones/carriers transmitted. Observe that PhaseU, the proposed schemes with limited polarization diversity, and that with full polarization diversity and EQU/MNV weighting, benefit little from the frequency diversity induced by the multi-tone transmissions. Meanwhile, the achievable AER under the full polarization diversity configuration with NVP weighting decreases exponentially with the number of signals transmitted.
\section{Conclusion}\label{conclusions}
In this work, we proposed a threshold-based LOS/NLOS classifier exploiting weighted differential (over different polarization configurations) carrier phase   measurements in a single mm-wave link. The proposed mechanism can work both in full and limited polarization diversity systems, and we developed a framework for assessing its performance. We conducted extensive simulations and showed that the performance of the proposed classifier, specified in terms of average error rate, depends critically on the potential reflecting materials in NLOS scenarios. For instance, the classifier is far more efficient in NLOS scenarios with wooden reflectors than in those with metallic reflectors. We illustrated the enormous performance gains that can be attained from exploiting full polarization diversity, properly weighting the differential carrier phase  measurements,  and using multi-tone transmissions. Indeed, we showed that our proposed approach exploiting full polarization diversity and noise variance -proportional weighting can significantly outperform the LOS/NLOS classifier proposed in \cite{Chenshu.2015}, which also exploits phase measurements, specially given a relatively high path power gain and/or the use of multiple tones. Finally, we showed that the optimum decision threshold remains approximately the same independently of the material of potential NLOS reflectors, while it decreases following a power-law as the average path/link power gain increases.

\bibliographystyle{IEEEtran}
\bibliography{IEEEabrv,references}
\end{document}